\documentclass[%
 reprint,
footinbib,
 amsmath,amssymb,
 aps,
prb,
]{revtex4-1}
\usepackage{graphicx}
\usepackage{color}
\usepackage{dcolumn}% Align table columns on decimal point
\usepackage{accents}% bold math
\definecolor{dgreen}{RGB}{00, 120, 00}
\definecolor{dcyan}{RGB}{ 30, 150, 170}
\usepackage[bookmarks=false]{hyperref}
\hypersetup{ colorlinks=true, linkcolor=blue, urlcolor=blue, citecolor=blue, }
\usepackage{booktabs}% Include figure files

\usepackage{ulem}

\begin{document}
\preprint{APS/123-QED}

\title{Local density of states in two-dimensional topological
superconductors under a magnetic field: signature of an exterior Majorana
bound state}

\author{Shu-Ichiro Suzuki}
\author{Yuki Kawaguchi}
\author{Yukio Tanaka}%

\affiliation{Department of Applied Physics, Nagoya University, Nagoya,
464-8603, Japan}

\date{\today}

\begin{abstract} 
We study quasiparticle states	on a surface of a
topological insulator (TI) with proximity-induced superconductivity under
an external magnetic field. An applied magnetic field creates two Majorana
bound states: a vortex Majorana state localized inside a vortex core and an
exterior Majorana state localized along a circle centered at the vortex
core. 
We calculate the spin-resolved local density of states (LDOS) and
demonstrate that the shrinking of the radius of the exterior Majorana
state, predicted in Ref.~[R.~S.~Akzyanov {\it et al.}, Phys.~Rev.~B {\bf
94}, 125428 (2016)], under a strong magnetic field can be seen in LDOS
without smeared out by non-zero-energy states. The spin-resolved LDOS
further reveals that the spin of the exterior Majorana state is strongly
polarized.  {Accordingly,} the induced odd-frequency spin-triplet pairs are
{found to be} spin-polarized as well.  In order to detect the exterior
Majorana states, however, the Fermi energy should be closed to the Dirac
point to avoid contributions from continuum levels. 
We also study a different two-dimensional topological-superconducting
system where a two-dimensional electron gas with the spin-orbit coupling is
sandwiched between an $s$-wave superconductor and a ferromagnetic
insulator.  We show that the radius of an exterior Majorana state can be
tuned by an applied magnetic field. However, on the contrary to the results
at a TI surface, neither the exterior Majorana state nor the induced
odd-frequency spin-triplet pairs are spin-polarized.  We conclude that the
spin-polarization of the Majorana state is attributed to the spin-polarized
Landau level which is characteristic for systems with the Dirac-like
dispersion. 
\end{abstract}

\pacs{???}% PACS, the Physics and Astronomy
                             % Classification Scheme.
\maketitle

%%%%%%%%%%%%%%%%%%%%%%%%%%%%%%%%%%%%%%%%%%%%%%%%%%%%%%%%%%%%%%%%%%%%%%%%%%%
%%%%%%%%%%%%%%%%%%%%%%%%%%%%%%%%%%%%%%%%%%%%%%%%%%%%%%%%%%%%%%%%%%%%%%%%%%%

\section{\label{sec:introduction} Introduction}

The existence of edge states is one of the most characteristic properties
of topological phases.  The edge states stem from the topology of bulk
wavefunctions, and have been actively studied in this decade
\cite{qi11,tanaka12,Beenakker13,Sato_review_2017}. In particular, exploring
Majorana bound states (MBSs) in topological superconductors (TSCs) is of great
worth \cite{Read, Ivanov01, Beenakker_2008_PRL, wilczek09, Law2009,
alicea11, Flensberg2012, alicea12, Beenakker13, Majo_review_2015,
Sato_review_2016, Flensberg_2017_PRL, Cayao_2017_PRB, lutchyn_review_2017,
Majo_review_2017}, since topologically protected quantum computation can be
implemented by braiding MBSs \cite{Kitaev03,Nayak}.  The presence of MBSs in
condensed matters has first been proposed in spinless $p$-wave
superconductors (SCs) \cite{Read, Kitaev01}. {Majorana bound
states have been considered to be
observed by means of current fluctuations \cite{Beenakker_2008_PRL} and
tunneling spectroscopy \cite{Law2009, Flensberg_2017_PRL, Cayao_2017_PRB}.}
Recently, more accessible
experimental setups have been proposed in various hybrid systems where a
spin-singlet $s$-wave SC is proximity-coupled to a low-dimensional
semiconductor with a strong spin-orbit coupling \cite{STF09, oreg10,
lutchyn10, SAU,ALICEA, Nakosai1, Yamakage, Bjornson_PRB_2013}.  Several
experimental studies found the evidences of MBSs \cite{Mourik, Deng,
Rokhinson, Das, yazdaniex, NW-exp2015, NW-exp2016, Majo_chen2016, Deng02,
Klinovaja_2016_exp,Marcus_PRL_2017,Suominen, Majo_exp_PRX2017,
Majo_Sestoft_2017, Majo_exp_gul_2018}. 

%%%%%%%%%%%%%%%%%%%%%%%%%%%%%%%( Schematic Figure )%%%%%%%%%%%%%%%%%%%%%%%
\begin{figure}[b]
	\begin{flushright}
		\includegraphics[width=0.46\textwidth]{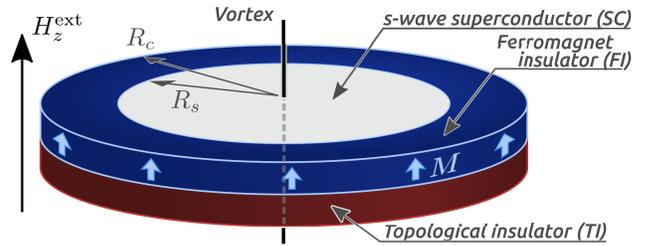}
		\caption{Schematic of superconductor (SC)/topological insulator (TI)
		hybrid system.  A superconducting island with a single vortex is
		fabricated on the top of a TI.  The SC is surrounded by a ferromagnetic
		insulator (FI) to bound the quasiparticles.  Therefore, the
		magnetization of the FI is perpendicular to the TI surface and its
		magnitude is much larger than the chemical potential and the pair
		potential.}
	\label{fig:sche}
	\end{flushright}
\end{figure}
%%%%%%%%%%%%%%%%%%%%%%%%%%%%%%%( Schematic Figure )%%%%%%%%%%%%%%%%%%%%%%

%%%%%%%%%%%%%%%%%%%%%%%%%%%%%%%%%%%%%%%%%%
% Topological insulator junctions
%%%%%%%%%%%%%%%%%%%%%%%%%%%%%%%%%%%%%%%%%%%
Among these hybrid systems, SCs on a three-dimensional topological
insulator (TI) are promising systems to observe and manipulate MBSs
\cite{FK08,FK09, TYN09,ANB09, SauTI, Ioselevich, veldhorst12, williams12}.
In such systems, a MBS appears as a localized state at a vortex core
\cite{Chamon, Cheng, Rakhmanov, TKawakami}. Simultaneously, another MBS
should appear somewhere in the system because MBSs can appear only in pairs
\cite{Flensberg2012}.  Recently, R.~S.~Akzyanov {\it et al.} have shown the
existence of exterior MBSs which are not always pinned by an interface nor a
sample edge \cite{Akzyanov15, Akzyanov16}.  Focusing only on the
zero-energy solutions, they have concluded that the position of the
exterior MBS can be controlled by an applied magnetic field
\cite{Akzyanov16}.  From experimental side, on the other hand, the local
conductance measurements by STS has become accessible in this hybrid system
\cite{Jin-Peng}.  Therefore, the study on the quasiparticle energy
spectrum, including the exterior MBS, non-zero-energy Andreev bound states,
and the continuum levels, is indispensable to observing MBSs and realizing
the braiding MBSs. {Motivated by these works, in this paper, we
investigate how one can identify the existence of the exterior MBS and its
position shift dependent on an applied magnetic field from the local
density of state (LDOS) experiment.}

%%%%%%%%%%%%%%%%%%%%%%%%%%%%%%%%%%%%%%%%%%%
% Odd-frequency
%%%%%%%%%%%%%%%%%%%%%%%%%%%%%%%%%%%%%%%%%%%

Another aspect of physics of MBSs is the appearance of odd-frequency Cooper
pairs. 
{ 
In general, Cooper pairs are classified into several symmetry classes by
focusing on the symmetry of the anomalous Green's function.  Most
theoretical papers have focused only on the equal-time pairing, of which
the anomalous Green's functions are even-function of the relative time of
two fermions. However, the anomalous Green's function depends generally on
the relative time (i.e., frequency), and can be an odd function of the
frequency. Such odd-frequency Cooper pairs are known to be locally induced
by spatial inhomogeneity as subdominant components \cite{odd1, odd3, odd3b,
	Asano_PRL_2007, Eschrig_Nat_2007, Yokoyama_2008_PRB, Tanuma_PRL_2009,
	Asano_PRL_2011, Daino_PRB_2012, Asano2013, Higashitani_PRL_2013, SIS1,
SIS3, Alidoust_2017_PRB, Kashuba_PRB_2017, Linder_Review} }.  Majorana
bound states are equivalent to zero-energy Andreev bound states
\cite{tanaka12}, which always always accompany subdominant odd-frequency
Cooper pairs. 
So far, the relation among odd-frequency pairing, MBSs, and TSCs have been
studied in a number of works \cite{Asano2013, Stanev, Balatsky1, SIS2,
Sau2015, Ebisu, Bjornson_PRB_2015,Ebisu2, Ebisu3, Cayao, Kuzmanovski}.  The
features of MBSs can be elucidated by focusing on the induced odd-frequency
pairings.  Therefore, it is worth investigating which type of Cooper pairs
is induced by the novel exterior MBS.

%%%%%%%%%%%%%%%%%%%%%%%%%%%%%%%%%%%%%%%
% Exterior Majoranay
%%%%%%%%%%%%%%%%%%%%%%%%%%%%%%%%%%%%%
In this paper, we study quasiparticle states on a TI surface where $s$-wave
superconductivity is proximity-induced, which we refer to as the SC/TI
hybrid system, and investigate how the LDOS is changed under an external
magnetic field normal to the TI surface.  Here, we put a vortex state of an
even-frequency spin-singlet $s$-wave SC on the TI so that the vortex MBS and
the exterior MBS arise.  From the calculated LDOS, we numerically confirm
that the wavefunction of the exterior MBS is localized on a circle around
the vortex and the radius shrinks as the external magnetic field increases.
In order to detect the exterior MBS from the LDOS, however, the Fermi energy
should be set close to the Dirac point to avoid contributions from
continuum levels. 
We also calculate the spin-resolved LDOS, and show that the exterior
MBS is strongly spin-polarized. The direction of the spin
polarization is determined by the direction of the external field. This
property is related to the lowest ``relativistic'' Landau level\cite{LL1,
LL2, LL3}, which is specific to systems with the Dirac-like dispersion such
as TI surfaces. Accordingly, we find that odd-frequency spin-polarized
$s$-wave Cooper pairs are accompanied by the exterior MBS. 

We also study a different two-dimensional TSC known as the Rashba SC: a
two-dimensional electron gas (2DEG) with a Rashba spin-orbit coupling (SOC)
sandwiched between an $s$-wave SC and a ferromagnetic insulator (FI).  We
show that the position of the exterior MBS is tunable by an applied magnetic
field as seen in SC/TI hybrid systems. 
{
However, contrary to the results in
SC/TI hybrid systems, the exterior MBS is not spin polarized.  This
difference can be well interpreted by comparing the spin structure of the
$n=0$ Landau level.  Reflecting that the $n=0$ Landau level is not
spin-polarized in a system with a conventional parabolic dispersion, the
exterior MBS is not spin polarized in the Rashba SC. 
}

%%%%%%%%%%%%%%%%%%%%%%%%%%%%%%%
The organization of this paper is as follows.  In Sec.~\ref{sec:BdG}, we
present the Bogoliubov-de Gennes (BdG) Hamiltonian for the SC/TI hybrid
system (Sec.~\ref{sec:BdG_SCTI}) and the Rashba SC (Sec.~\ref{sec:BdG_RSC})
and the formulation to obtain the energy spectrum, LDOS, and pair
amplitudes (Sec.~\ref{sec:BdG_eigenfunction}).  In Sec.~\ref{sec:results},
we numerically calculate the energy spectrum, LDOS, and odd-frequency
pairing both for SC/TI hybrid systems and Rashba SCs.
Section~\ref{sec:conclusion} summarizes our results.  

%%%%%%%%%%%%%%%%%%%%%%%%%%%%%%%%%%%%%%%%%%%%%%%%%%%%%%%%%%%%%%%%%%%%%%%%%%%
%%%%%%%%%%%%%%%%%%%%%%%%%%%%%%%%%%%%%%%%%%%%%%%%%%%%%%%%%%%%%%%%%%%%%%%%%%%
%%%%%%%%%%%%%%%%%%%%%%%%%%%%%%%%%%%%%%%%%%%%%%%%%%%%%%%%%%%%%%%%%%%%%%%%%%%

\section{Bogoliubov-de Gennes formalism}
\label{sec:BdG}

\subsection{Superconductor/topological insulator hybrid system}
\label{sec:BdG_SCTI}
We consider a surface of a three-dimensional TI on which a superconducting
island is fabricated as shown in Fig.~\ref{fig:sche}, where we use a
type-II SC and a vortex is assumed to be located at the center of the SC.
The SC is surrounded by an FI to bound quasiparticles at the SC/TI
interface: The magnetization of the FI is directed perpendicular to the TI
surface and its amplitude is much larger than the chemical potential and
the pair potential so that {a large} energy gap opens at the FI/TI
interface. The radii of the SC and the system are denoted by $R_s$ and
$R_c$, respectively. 
{ The existence of the FI is not essential when an exterior MBS is
localized inside the superconducting region under a strong magnetic field.
The FI is needed to localize an exterior MBS at the SC/FI boundary under a
small magnetic field.  Because a TI surface is already an interface between
a 3DTI and a vacuum, we cannot terminate the metallic TI surface by a
vacuum but need to introduce another type of insulating state by putting an
FI on it. From the point of view of numerical calculations, the existence
of the FI makes the boundary condition simple so that quasi-particle
wavefunctions vanish at the boundary of the system.}

When the superconductivity is induced by attaching a SC on the top of the
TI, the system can be described by the Hamiltonian as
\begin{align}
  &\mathcal{H} 
	= \frac{1}{2} \iint ~
	\Psi^\dagger (\boldsymbol{r})
	\check{H}_{B}(\boldsymbol{r}, \boldsymbol{r}') 
	\Psi         (\boldsymbol{r}')
	\, d\boldsymbol{r} d\boldsymbol{r}', \\[0mm]
	&\Psi           (\boldsymbol{r}) = 
	[\psi_{\uparrow}           (\boldsymbol{r}) ~
	 \psi_{\downarrow}         (\boldsymbol{r}) ~
	 \psi_{\uparrow}  ^\dagger (\boldsymbol{r}) ~
   \psi_{\downarrow}^\dagger (\boldsymbol{r}) ]^{\mathrm{T}}, 
\end{align}
with the Bogoliubov-de Gennes (BdG) Hamiltonian 
\begin{align}
	&\check{H}_{B}(\boldsymbol{r}, \boldsymbol{r}') 
	= 
	\left[
		\begin{array}{cc}
			\delta( \boldsymbol{r}- \boldsymbol{r}')
			 \hat{h     }  ( \boldsymbol{r})                  &
			 \hat{\Delta}  ( \boldsymbol{r}, \boldsymbol{r}') \\
			-\hat{\Delta}^*( \boldsymbol{r}, \boldsymbol{r}') &
			-\delta( \boldsymbol{r}- \boldsymbol{r}')
			 \hat{h     }^*( \boldsymbol{r}) \\
	\end{array}\right], 
	%%
	%&= \delta(\boldsymbol{r}-\boldsymbol{r}')\left[
	%	\begin{array}{cc}
	%		 \hat{H     }  ( \boldsymbol{r})                  &
	%		 {\Delta}  ( \boldsymbol{r}) i \sigma_2\\
	%		-{\Delta}^*( \boldsymbol{r}) i \sigma_2&
	%		-\hat{H     }^*( \boldsymbol{r}) \\
	%\end{array}\right]
\end{align}
where $\boldsymbol{r} = (x,y)$ and the symbol $\hat{\cdot}$
($\check{\cdot}$) represents a $2 \times 2$ ($4 \times 4$) matrix in the
spin (spin-Nambu) space.  The single-particle Hamiltonian $\hat{h}$ and the
pair potential $\hat{\Delta}$ can be written as 
\begin{align}
	&\hat{h}(\boldsymbol{r})
	= v_F \hat{\boldsymbol{\sigma}} \cdot 
	\left[{\boldsymbol{p}} 
	-\frac{e}{c} \boldsymbol{A}(\boldsymbol{r}) \right] 
	+ M(\boldsymbol{r}) \hat{\sigma}_3
	-\mu_F \hat{\sigma}_0, 
	\\
	&\hat{\Delta}  ( \boldsymbol{r}, \boldsymbol{r}') 
	= \delta(\boldsymbol{r}-\boldsymbol{r}') \Delta ( \boldsymbol{r} ) i\hat{\sigma}_2,
\end{align}
where $v_F$, $e<0$, $c$, $\boldsymbol{p} = -i \hbar \boldsymbol{\nabla}$,
$\hat{\sigma}_i$ ($i=0$-$3$), $\boldsymbol{A} (\boldsymbol{r})$, and
$\mu_F$ are the Fermi velocity, the charge of an electron, the speed of
light, the momentum operator, the Pauli matrices in the spin space, the
vector potential, and the chemical potential, respectively. We here assume
that the pair potential $\hat{\Delta}$ has the even-frequency spin-singlet
$s$-wave symmetry.  The magnetization of the FI and the amplitude of the
proximity-induced pair potential are denoted by $M(\boldsymbol{r})$ and
$\Delta(\boldsymbol{r})$, respectively. 

When the system is cylindrically symmetric, the Hamiltonian can further be
reduced by introducing the cylindrical coordinate;
$\boldsymbol{r}=(\rho,\phi)$.  We assume that the $\phi$ dependence of the
pair potential is written as $\Delta (\boldsymbol{r}) = \Delta(\rho) e^{-i
\ell \phi}$ with $\ell$ being the vorticity and that the magnetization
$M(\boldsymbol{r})$ does not depend on $\phi$.  Then, by expanding the
field operators as $\Psi(\phi,\rho) = \sum_{\mu} \Psi_{\mu}(\rho) e^{i \mu
\phi} / (2 \pi)^{1/2}$ with the basis 
\begin{align}
	\Psi_\mu(\rho) 
	= \left( \begin{array}{l}
%  \psi        _{\uparrow  , \mu+\ell  } e^{i  \ell    \phi} \\
%	\psi        _{\downarrow, \mu+\ell+1} e^{i (\ell+1) \phi} \\
%	\psi^\dagger_{\uparrow  ,-\mu-1     } e^{i          \phi} \\
%	\psi^\dagger_{\downarrow,-\mu       } \\
	\psi        _{\uparrow  , \mu}(\rho) e^{-i (\ell+1) \phi/2} \\[2mm] 
	\psi        _{\downarrow, \mu}(\rho) e^{-i (\ell-1) \phi/2} \\[2mm]
	\psi^\dagger_{\uparrow  , \mu}(\rho) e^{+i (\ell+1) \phi/2} \\[2mm]
	\psi^\dagger_{\downarrow, \mu}(\rho) e^{+i (\ell-1) \phi/2} \\
	\end{array} \right), 
	\label{eq:ft}
\end{align}
where $\mu$ is an integer (a half integer) for an odd (even) vorticity to
make $\Psi$ a single-valued function, the Hamiltonian is rewritten as 
\begin{gather}
  \mathcal{H} 
	= \frac{1}{2} \sum_{\mu} \int_{0}^{R_c}
  \Psi^\dagger_{\mu} (\rho) 
	\check{H}_{B, \mu}  ( \rho )
	\Psi_{\mu}   (\rho)
	~\rho d\rho, %\\
	\end{gather}
	\vspace{-3mm}
	\begin{gather}
	\check{H}_{B, \mu}  ( \rho )
	   = \left[
	  	\begin{array}{cc}
				\hat{h     }_{\mu}  ( \rho ) &
				 {\Delta}  ( \rho ) i \hat{\sigma}_2 \\[2mm]
	  		-{\Delta}^*( \rho ) i \hat{\sigma}_2 &
				-\hat{h     }_{-\mu}^*( \rho ) \\
		\end{array}\right].% \\[2mm]
\label{eq:H_Bmu}
\end{gather}
{
The $4 \times 4$ Hamiltonian $\check{H}_{B,\mu}$ preserves the
particle-hole symmetry 
\begin{align}
	\check{H}_{B,\mu} = -\check{\tau}_1 \check{H}^*_{B,-\mu} \check{\tau}_1, 
	\label{eq:sym}
\end{align}
}where $\check{\tau}_i$ ($i=1$-$3$) are the Pauli matrices in the Nambu
space.  The single-particle Hamiltonian $\hat{h     }_{\mu}  ( \rho )$ is
given by
\begin{widetext}
\begin{align}
	\hat{h}_{\mu}( \rho )
	 = 
  \left[ \begin{array}{cc}
			M(\rho)-\mu_F &
			\cfrac{\hbar v_F}{i} 
      \left( \partial_\rho +\cfrac{2\mu-\ell+1}{2\rho} 
             +\tilde{A}_\phi(\rho) \right) \\
			\cfrac{\hbar v_F}{i} 
      \left( \partial_\rho -\cfrac{2\mu-\ell-1}{2\rho} 
      -\tilde{A}_\phi(\rho) \right) &
			-M(\rho)-\mu_F \\
	 \end{array} \right].
\end{align}
\end{widetext}
Here, we chose the gauge $\boldsymbol{A}(\boldsymbol{r}) = A_\phi(\rho)
\boldsymbol{e}_\phi$ with $\boldsymbol{e}_\phi$ being a unit vector along
the azimuthal direction and defined $\tilde{A}_\phi = |e|A_\phi / \hbar c$.
The vector potential is due to an external magnetic field in addition to
the magnetic flux of a vortex. 
Throughout this paper, we assume that the thickness of the SC along the $z$
direction is much thinner than its magnetic penetration depth so that the
magnetic field is spatially homogeneous on the TI surface. The magnitude of
an external magnetic field is expressed in terms of the magnetic length
$\ell_B = ( \hbar c / |e H_z^{\mathrm{ext}}|)^{1/2}$,
which should be larger than the coherence length $\xi_{\mathrm{SC}}=\hbar
v_F /\Delta_{\mathrm{SC}}$ of the bulk SC with pair potential
$\Delta_{\mathrm{SC}}$: otherwise, the magnetic field destroys the
superconductivity.  On the other hand, $\xi_{\mathrm{SC}}$ is much smaller
than the coherence length $\xi_0 = \hbar v_F /\Delta_0$ of the
proximity-induced pair potential $\Delta_0$ on the TI surface (i.e.,
$\Delta_0\ll\Delta_{\mathrm{SC}}$).  In the following calculations, we
consider the case for $\xi_{\mathrm{SC}}\ll \xi_0 < \ell_B$. 

To describe the $\rho$ dependence around the vortex core, we assume 
\begin{align}
  \Delta(\rho) 
	&= \Theta( R_s - \rho ) \, \Delta_0 \tanh \left( \rho / \xi_0 \right),
\label{eq:Delta_rho}
\end{align}
where $R_s$ is the radius of the SC, $\Delta_0$ is the amplitude of the
proximity-induced pair potential in the absence of a vortex, and
$\Theta(\rho)$ is the Heaviside step function.  The radial profile of the
magnetization is given by $M(\rho) = M_0 \Theta(\rho-R_s)\Theta(R_c-\rho)$
with $R_c$ being the outer radius of the FI. 

%%%%%%%%%%%%%%%%%%%%%%%%%%%%%%%%%%%%%%%%%%%%%%%%%%%%%%%%%%%%%%%%%%%%%%%%%%%
%%%%%%%%%%%%%%%%%%%%%%%%%%%%%%%%%%%%%%%%%%%%%%%%%%%%%%%%%%%%%%%%%%%%%%%%%%%
%%%%%%%%%%%%%%%%%%%%%%%%%%%%%%%%%%%%%%%%%%%%%%%%%%%%%%%%%%%%%%%%%%%%%%%%%%%

\subsection{Rashba superconductor}
\label{sec:BdG_RSC}
\begin{figure}[tb]
	\begin{flushright}
		\hspace{0.0cm}\includegraphics[width=0.46\textwidth]{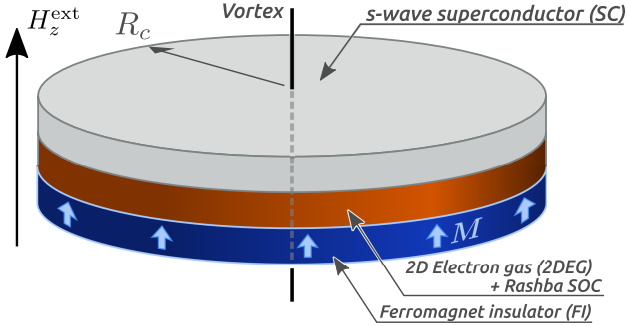}
		\caption{Schematic of Rashba SC system. The two-dimensional electron gas (2DEG) is
		sandwiched between an FI and an $s$-wave SC with a vortex. The magnetization of the FI
	is normal to the 2DEG.}
	\label{fig:sche2}
	\end{flushright}
\end{figure}

We also consider a 2DEG with a Rashba SOC. The 2DEG is sandwiched between
an even-frequency spin-singlet $s$-wave SC and an FI, where a
superconducting vortex is located at the center of the SC and the
magnetization of the FI is normal to the 2DEG as shown in
Fig.~\ref{fig:sche2}. In this paper, we refer to this system as the Rashba
SC. {Both of the two systems (Figs.~\ref{fig:sche} and
\ref{fig:sche2}) are two-dimensional islands of a topological
superconductor surrounded by an insulating state even though the geometries
look different. The appearances are different from each other because FIs
play different roles; an FI is introduced to make an insulating region in
the SC/TI hybrid system, whereas it is introduced to make a
superconductivity topological in the Rashba SC.}

The single-particle Hamiltonian for the Rashba SC can be described
\cite{STF09, lutchyn10, SAU,ALICEA, Nakosai1, Yamakage} as 
\begin{align}
	\hat{h} 
	&= \left[ \frac{\tilde{\boldsymbol{p}}^2}{2m_0 }  - \mu_F \right] \hat{\sigma}_0
	+ \lambda \boldsymbol{e}_z \cdot 
	  \left[ \hat{\boldsymbol{\sigma}} \times \tilde{\boldsymbol{p}} \right]
	+ \boldsymbol{M} \cdot \hat{\boldsymbol{\sigma}}, 
\end{align}
where $\lambda$ is the strength of the {SOC}, $\tilde{\boldsymbol{p}} =
\boldsymbol{p}-(e/c)\boldsymbol{A}$, $\boldsymbol{e}_z$ is the unit vector
in the $z$-direction, $m_0$ is the mass of an electron. In the Cartesian
coordinate, $\hat{h}$ can be written as 
\begin{align}
	\hat{h}(x,y) = 
	\left[ \begin{array}{cc}
			\xi_k + M_0 & 
			 \lambda \Big(  \tilde{\partial}_x -i \tilde{\partial}_y \Big) \\
			 \lambda \Big(- \tilde{\partial}_x -i \tilde{\partial}_y \Big) & 
			\xi_k - M_0 \\
	\end{array} \right], 
	\label{}
\end{align}
where we have used $\boldsymbol{M} = M_0 \boldsymbol{e}_z$ with $M_0$ being
a constant, $\xi_k = (-\hbar^2/2m_0)( \tilde{\partial}_x^2 +
\tilde{\partial}_y^2 ) - \mu_F$ and $\tilde{\partial}_{x(y)} =
\partial_{x(y)} +i |e| A_{x(y)}/\hbar c $.  The system becomes
topologically non-trivial (i.e., the 2DEG becomes a TSC) when the relation
$M_0^2 > {\mu_F^2 + \Delta_0^2}$ is satisfied \cite{STF09, oreg10,
lutchyn10, SAU, Yamakage}. 

Similarly to the previous section, the Hamiltonian can be simplified when
the system has the rotational symmetry.  Assuming $\Delta( \rho, \phi) =
\Delta(\rho) e^{-i \ell \phi} = \Delta_0 \tanh (\rho / \xi_0) e^{-i \ell
\phi}$ and expanding the field operators as Eq.~\eqref{eq:ft}, the
Hamiltonian can be written as 
\begin{align}
  &\mathcal{H} 
	= \frac{1}{2} \sum_{\mu} \int_{0}^{R_c}
  \Psi^\dagger_{\mu} (\rho) 
	\check{H}_{B, \mu}  ( \rho )
	\Psi_{\mu}   (\rho)
	~\rho d\rho, \\[1mm]
	&\check{H}_{B, \mu}  ( \rho )
	   = \left[
	  	\begin{array}{cc}
				\hat{h}_{\mu}  ( \rho ) &
				 {\Delta}  ( \rho ) i \hat{\sigma}_2 \\[2mm]
	  		-{\Delta}^*( \rho ) i \hat{\sigma}_2 &
				-\hat{h}_{-\mu}^*( \rho ) \\
		\end{array}\right],
		\label{eq:HB_RM}
\end{align} 
where the diagonal part is given by 
\begin{widetext}
\begin{align}
	\hat{h}_\mu(\rho) = 
	\left[ \begin{array}{cc}
			\xi_{\mu} + M_0 & 
			\lambda \left(  \partial_\rho + \cfrac{2\mu-\ell+1}{2\rho} +\tilde{A}_\phi \right) \\
			\lambda \left( -\partial_\rho + \cfrac{2\mu-\ell-1}{2\rho} +\tilde{A}_\phi \right) &
			\xi_{\mu+1} - M_0 \\
	\end{array} \right], 
\end{align}
with
\begin{align}
	\xi_\mu = - \frac{\hbar^2}{2m_0} 
	\left[ 
	   \frac{1}{\rho} \frac{\partial}{\partial \rho}
	    \left( \rho \frac{\partial}{\partial \rho} \right)
	    - \frac{(2\mu -\ell -1)^2 }{4\rho^2} 
			+ \frac{\tilde{A}_\phi(2\mu -\ell -1) }{2 \rho} 
			- |\tilde{A}_\phi|^2
	  \right] - \mu_F, 
\end{align}
\end{widetext}
being the kinetic energy in the cylindrical coordinate.  The definition of
$\tilde{A}_\phi$ is the same as that in Sec.~\ref{sec:BdG_SCTI}.
$\check{H}_{B, \mu}$ in Eq.~\eqref{eq:HB_RM} also preserves the
particle-hole symmetry described in Eq.~(\ref{eq:sym}).

%%%%%%%%%%%%%%%%%%%%%%%%%%%%%%%( Schematic Figure )%%%%%%%%%%%%%%%%%%%%%%%%%%%%%%%%%%%%%%%
%%%%%%%%%%%%%%%%%%%%%%%%%%%%%%%( Schematic Figure )%%%%%%%%%%%%%%%%%%%%%%%%%%%%%%%%%%%%%%%

%%%%%%%%%%%%%%%%%%%%%%%%%%%%%%%%%%%%%%%%%%%%%%%%%%%%%%%%%%%%%%%%%%%%%%%%%%%%%%%%%%%%%%%%%%
\subsection{Local density of states and pair amplitudes}
\label{sec:BdG_eigenfunction}
The local density of states (LDOS) and the pair amplitudes can be calculated from the
quasiparticle eigenfunctions. We numerically solve the BdG equation
\begin{align}
	H_{B,\mu} \Phi_{\mu,\nu} = E_{\mu,\nu} \Phi_{\mu,\nu}, 
\end{align}
for each $\mu$, and obtain the eigenfunctions 
$\Phi_{\mu,\nu} = \left[~
u_{  \uparrow,\mu,\nu}~ ~
u_{\downarrow,\mu,\nu}~ ~ 
v_{  \uparrow,\mu,\nu}~ ~ 
v_{\downarrow,\mu,\nu}~ \right]^{\mathrm{T}}$ 
where $u_{s,\mu,\nu}$ and $v_{s,\mu,\nu}$ are the wavefunctions of
quasiparticles in spin $s=$~$\uparrow$ and $\downarrow$ states and $\mu$ and
$\nu$ are the indices that specify the $\phi$ dependence of the wavefunction and
the eigenvalue, respectively. In the numerical simulations, we expand the
eigenfunction in terms of the Bessel functions and numerically diagonalize the
BdG Hamiltonian. The details are described in Appendices \ref{sec:app_A} (SC/TI
hybrid system) and \ref{sec:app_B} (Rashba SC).

Using the eigenfunctions, the LDOS is expressed as 
\begin{align}
	&N(\rho, E) 
	= \sum_{s=\uparrow,\downarrow} N_s(\rho, E), \\
	&N_s(\rho, E) 
	= \sum_{\mu} \sum_{E_{\mu,\nu} \geqslant 0}
	\left[  |u_{s,\mu,\nu}(\rho)|^2\, \eta( E - E_{\mu,\nu}) \right. \notag \\[-2mm]
& \hspace{33mm}\left. +|v_{s,\mu,\nu}(\rho)|^2\, \eta( E + E_{\mu,\nu}) \right]. 
\end{align}
In this paper, we employ the thermal-smearing function $\eta (E) = -
(\partial/\partial E) f(E,T)$, where $f(E,T)$ is the Fermi-Dirac distribution
function.  The temperature is set to ${T=0.1 T_c}$ for the SC/TI hybrid systems
and ${T=0.01 T_c}$ for the Rashba SCs, where the critical temperature $T_c$ is
obtained from the relation between $T_c$ and $\Delta_0$ in an $s$-wave singlet
superconductor: $T_c = \Delta_0 e^{\gamma} / \pi$, where $\gamma$ is Euler's
gamma.

We can obtain the anomalous Green's function from the eigenfunctions as well.
Throughout this paper, we refer to the anomalous Green's function as a pair
amplitude. In general, for a given set of eigenfunctions
$[~u_{\uparrow,\nu}~ ~u_{\downarrow,\nu}~ ~v_{\uparrow,\nu}~
~v_{\downarrow,\nu}~]^{\mathrm{T}}$
and associated eigenvalues $E_\nu$, the anomalous Green's function is given by
\begin{align}
 &\mathfrak{F}_{\sigma_1 \sigma_2} (i \omega_n; \boldsymbol{r}_1,\boldsymbol{r}_2) \notag \\
 & \hspace{03mm}=
	\sum_{E_\nu \geqslant 0} \Bigg[
		 \frac{u  _{\sigma_1,\nu} (\boldsymbol{r}_1) \, v^*_{\sigma_2,\nu}
		 (\boldsymbol{r}_2)} {i \omega_n -E_\nu} \notag \\[-2mm]
		 & \hspace{20mm}
		+\frac{v^*_{\sigma_1,\nu} (\boldsymbol{r}_1) \, u  _{\sigma_2,\nu} 
		(\boldsymbol{r}_2)} {i \omega_n +E_\nu}
 \Bigg], \label{fig:gfa-o}
\end{align}
where $\omega_n = (2n+1) \pi T$ is the Matsubara frequency with $n$ being an integer. 
In the present case, by recovering the $\phi$ dependence of the quasiparticle
wavefunctions using Eq.~(\ref{eq:ft}), the anomalous Green's function can be expressed as 
\begin{align}
	&\mathfrak{F}_{s_1 s_2} ( i\omega_n; \boldsymbol{r}_1, \boldsymbol{r}_2 ) 
	= \sum_{\mu} \mathfrak{F}_{\mu, s_1 s_2} ( i\omega_n; \boldsymbol{r}_1,
	\boldsymbol{r}_2 ), \\
% ( i\omega_n; \rho_1, \phi_1, \rho_2, \phi_2 ) \\
  & \mathfrak{F}_{\mu, s_1 s_2} ( i\omega_n; \boldsymbol{r}_1, \boldsymbol{r}_2 ) \notag \\
	& \hspace{01mm} =
	\sum_{E_\nu \geqslant 0} \Big[
		 \frac{u  _{s_1,\mu,\nu} (\rho_1) \, v^*_{s_2,\mu,\nu} (\rho_2)} 
		 {i \omega_n -E_{\mu,\nu}} e^{+i \mu \phi_r} \notag \\
		 & \hspace{18mm}+\frac{v^*_{s_1,\mu,\nu} (\rho_1) \, u  _{s_2,\mu,\nu} (\rho_2)} 
	   {i \omega_n +E_{\mu,\nu}} e^{-i \mu \phi_r}  \Big] \notag \\[2mm]
		 & \hspace{30mm} \times e^{-i \ell \phi_c}  e^{-i(X_{s_1} \phi_1 + X_{s_2} \phi_2)/2}, 
		 \label{fig:gfn-o} 
\end{align}
with $\phi_c = (\phi_1 + \phi_2)/2$, $\phi_r = \phi_1 - \phi_2$, and $X_{s}=1(-1)$ for $s=$
$\uparrow$ ($\downarrow$).  In this paper we focus on the $s$-wave component (i.e.,
on-site correlation function).  We can obtain the even- and odd-frequency components from the
following relation:
\begin{align}
	&{F}_{\mu, s_1 s_2}^{\mathrm{Even (Odd)}} ( i\omega_n; \rho_1, \phi_1 ) \notag \\
	&\hspace{04mm}=\frac{1}{2} \Bigg[ 
		\mathfrak{F}_{\mu, s_1 s_2} ( i\omega_n; \rho_1, \phi_1, \rho_1, \phi_1 ) \notag \\[-3mm]
	& \hspace{15mm} +(-)
	\mathfrak{F}_{\mu, s_1 s_2} (-i\omega_n; \rho_1, \phi_1, \rho_1, \phi_1 )
  \Bigg]. 
\end{align}

%%%%%%%%%%%%%%%%%%%%%%%%%%%%%%%%%%%%%%%%%%%%%%%%%%%%%%%%%%%%%%%%%%%%%%%%%%%%%%%%%%%%%%%%%%%

\section{Results}
\label{sec:results}
\subsection{Superconductor/topological insulator hybrid system}
%%%%%%%%%%%%%%%%%%%%%%%%%%%%%%%( Figure LDOS )%%%%%%%%%%%%%%%%%%%%%%%%%%%%%
\begin{figure}[tb]
	\begin{center}
		\hspace{0.0cm}\includegraphics[width=0.42\textwidth]{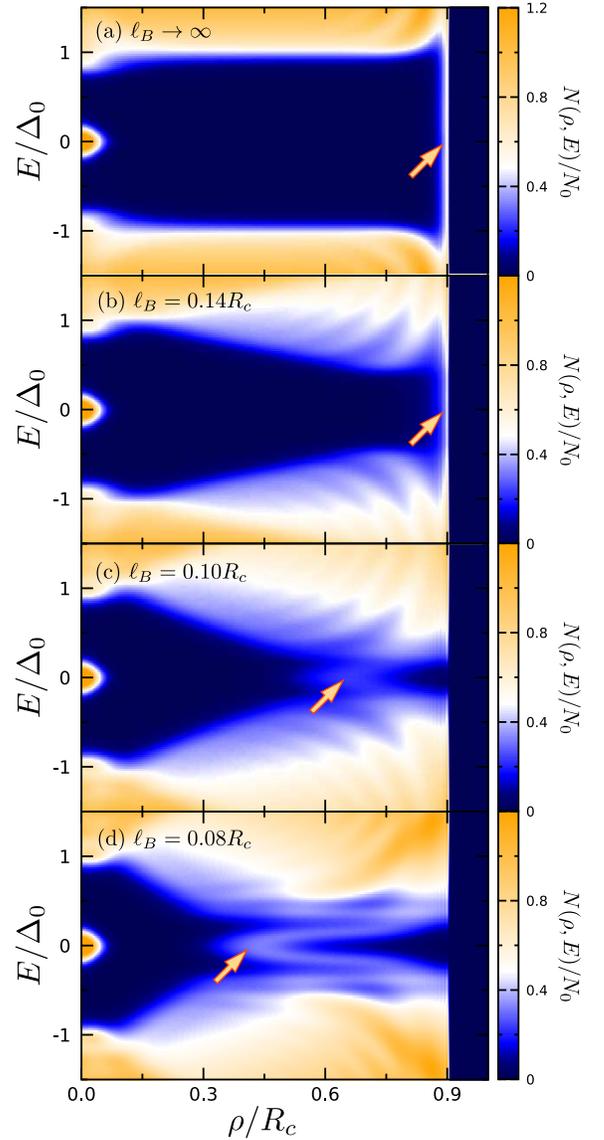}
		\caption{Local density of states (LDOS) in the SC/TI hybrid system. The
			vortex is located at the center of the superconducting island $\rho
			=0$.  The peaks at $\rho = E = 0$ are from the vortex Majorana bound
			states (MBS).  { The exterior MBS changes its position
				depending on an external magnetic field (indicated with the
				arrows). The V-shape peaks near the arrows in (c) and (d) are the
				characteristic structure for the exterior and chiral MBSs.  } The
				LDOS are normalized by the value at $\rho = 0$ and $E=1.5 \Delta_0$
				in the absence of the proximity-induced pair potential (i.e.,
				$\Delta_0 = 0$). The parameters are set as $\mu_F = 0.2\Delta_0$,
			$R_c = 30 \xi_0$, $R_s = 0.9R_c$, $\ell=1$, and $M_0 = 10\Delta_0$.  }
	\label{fig:ldos_fe02}
	\end{center}
\end{figure}

%%%%%%%%%%%%%%%%%%%%%%%%%%%%%%%( Schematic Figure )%%%%%%%%%%%%%%%%%%%%%%%%%

%%%%%%%%%%%%%%%%%%%%%%%%%%%%%%%%%%%%%%%%%%%%%%%%%%%%%%%%%%%%%%%%%%%%%%%%%%
%%%%%%%%%%%%%%%%%%%%%%%%%%%%%%%%%%%%%%%%%%%%%%%%%%%%%%%%%%%%%%%%%%%%%%%%%
\subsubsection{Local density of states}

We first show the LDOS with several choices of the magnetic length in
Fig.~\ref{fig:ldos_fe02}, where the chemical potential is set to $\mu_F =
0.2\Delta_0$.  The LDOS $N(\rho, E)$ in this system is normalized by $N_0 =
\bar{N}(\rho=0, E=1.5\Delta_0)$ for each chemical potential, where
$\bar{N}(\rho,E)$ is the LDOS of the pristine surface of the TI in the
absence of a proximity-induced pair potential nor an external magnetic
field.  In the absence of an external magnetic field (i.e., $\ell_B
\rightarrow \infty$), the proximity-induced pair potential opens a
spatially-homogeneous energy gap almost everywhere in the superconducting
region [Fig.~\ref{fig:ldos_fe02}(a)].  A vortex MBS appears at
the center of the vortex core at $E=0$.  The peaks at $\rho=0$ and
$E=\pm0.95\Delta_0$ are also the localized states at the vortex core. The
exterior of the superconducting region (i.e., $\rho > R_s$) is insulating
because of the magnetization of the FI. As a result, another MBS\cite{Klinovaja_Majo} appears at the SC/FI boundary (i.e., $\rho =
R_s$). 

With decreasing the magnetic length, the magnitude of the superconducting
gap near the SC/FI boundary decreases as shown in
Fig.~\ref{fig:ldos_fe02}(b).  When the magnetic length becomes shorter than
a certain value, the MBS at the SC/FI boundary moves
inside the superconducting region. In Figs.~\ref{fig:ldos_fe02}(c) and
\ref{fig:ldos_fe02}(d), for example, the zero-energy peaks are located at
$\rho \approx 0.7R_c$ and $0.4R_c$, respectively. The radius of the
exterior MBS $r^*$ becomes smaller for larger magnetic
field and is consistent with the relation $r^* = 2 \ell_B^2 / \xi_0 $
derived in Ref.~\onlinecite{Akzyanov16}.

In addition to the zero-energy states, there are also subgap ``edge''
states with non-zero eigenenergies. These states, known as chiral Majorana
states, stem from the solutions for $\mu \neq 0$ and exhibits linear
dispersion with respect to $\mu$ for small $\mu$ (see
Fig.~\ref{fig:mudep}).  They are localized at the SC/FI boundary when
$H_z^{\mathrm{ext}}=0$. When $H_z^{\mathrm{ext}}$ becomes stronger than a
certain value, they move inside the superconducting region as the exterior
Majorana zero-energy state does.  The radii of these states depend on their
eigenenergies: the higher-energy state, or equivalently the
larger-angular-momentum state, has the larger radius.  As a result, the
LDOS exhibits a characteristic V-shaped peak in the $\rho$\,-$E$ space as
shown in Figs.~\ref{fig:ldos_fe02}(c) and \ref{fig:ldos_fe02}(d).  The
chiral Majorana states exist even in the absence of a superconducting
vortex, although the exactly-zero-energy state arises only in the presence
of a vortex \cite{Law2009, alicea12}.  We observe numerically that the
chiral Majorana states without a vortex also move inside the
superconducting region under a strong external magnetic field.

\begin{figure}[tb]
	\begin{center}
		\hspace{0.0cm}\includegraphics[width=0.42\textwidth]{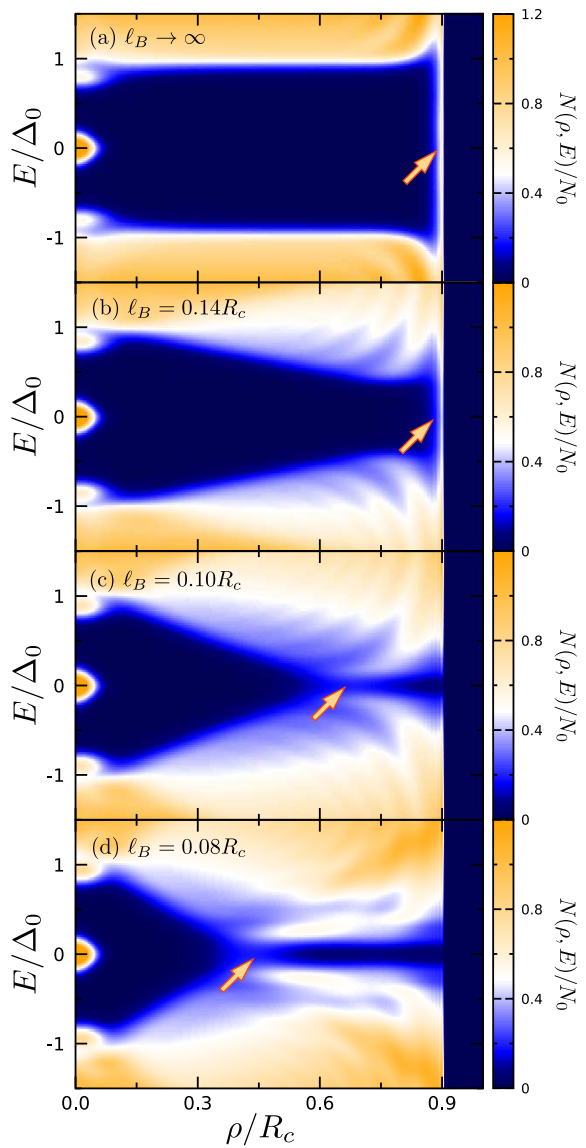}
		\caption{LDOS in the SC/TI hybrid system at $\mu_F = 0.4\Delta_0$. The other
		parameters are set as the same values as those used in
		Fig.~\ref{fig:ldos_fe02}. 
		{The arrows indicate the position of the exterior
		Majorana bound states. The characteristic V-shape is no longer clear.} }
	\label{fig:ldos_fe04}
	\end{center}
\end{figure}

We next show the LDOS for a larger chemical potential $\mu_F = 0.4\Delta_0$
(i.e., with a larger Fermi surface). In the absence of an external magnetic
field, the magnitude of the energy gap is equivalent to $\Delta_0$ except
for the vortex core and the boundary as shown in
Fig.~\ref{fig:ldos_fe04}(a).  When the magnetic field is applied, the
energy gap becomes smaller around the SC/FI boundary as shown in
Fig.~\ref{fig:ldos_fe04}(b).  At the magnetic length $\ell_B = 0.1R_c$ and
$0.08R_c$, $N(E=0,\rho)$ has a peak at $\rho \approx 0.7R_c$ and $0.4 R_c$,
respectively. However, these points are saddle points in the $\rho$-$E$
plane (i.e., these points are local minima along the $E$ direction).
Accordingly, the characteristic V-shape structure in the $\rho$\,-$E$
space, which can be seen in Fig.3(d), is segmented at $E=0$ in
Fig.~\ref{fig:ldos_fe04}(d). 

%%%%%%%%%%%%%%%%%%%%%%%%%%%%%%%( Schematic Figure )%%%%%%%%%%%%%%%%%%%%%%%%%%%%%%%%%%%%%%%
\begin{figure}[tb]
	\begin{center}
		\hspace{0.0cm}\includegraphics[width=0.44\textwidth]{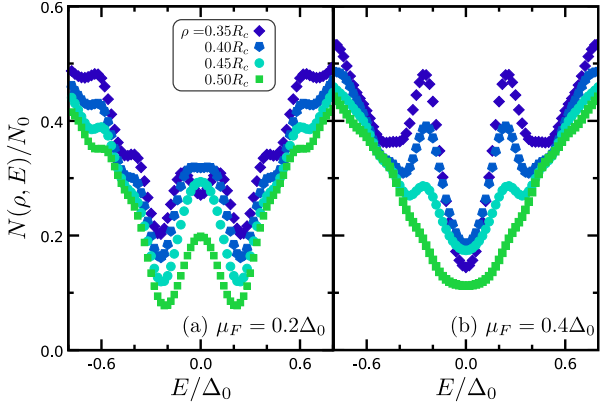}
		\caption{LDOS in the SC/TI hybrid system.  The same data as (a)
			Fig.~\ref{fig:ldos_fe02}(d) and (b) Fig.~\ref{fig:ldos_fe04}(d),
			which are respectively calculated for $\mu_F=0.2\Delta_0$ and
			$0.4\Delta_0$, are shown as a function of $E/\Delta_0$ at
			$\rho/R_c=0.35$, $0.40$, $0.45$, and $0.50$.  {When the
			chemical potential $\mu_F$ is close to the Dirac point (a), the
			exterior Majorana state can be observed as a zero-energy
			conductance peak. On the other hand, LDOS for a fixed $\rho$ has a
			minimum at $E=0$ when $\mu_F$ is not sufficiently small (b).} }
	\label{fig:ldos_cut}
	\end{center}
\end{figure}
%%%%%%%%%%%%%%%%%%%%%%%%%%%%%%%( Schematic Figure )%%%%%%%%%%%%%%%%%%%%%%%%

To understand the details of the LDOS for each chemical potential, we plot
the cross section of $N(\rho,E)$ for $\mu_F = 0.2\Delta_0$ and $0.4
\Delta_0$ in Figs.~\ref{fig:ldos_cut}(a) and \ref{fig:ldos_cut}(b),
respectively.  The radial coordinate is fixed at $\rho / R_c = 0.35, 0.40,
0.45$, and $0.50$ (i.e., around the zero-energy peak of the exterior
MBSs). When the chemical potential is sufficiently small, the
LDOS has a clear peak {at $E=0$} as shown in Fig.~\ref{fig:ldos_cut}(a).
However, when $\mu_F = 0.4 \Delta_0$, there is no zero-energy peak for
every $\rho$ even though there is an exterior MBS. 

Figure~\ref{fig:mudep} shows the angular-momentum dependences of the energy
eigenvalues at $\ell_B = 0.08 R_c$ and (a) $\mu_F / \Delta_0 =0.2$, (b)
$0.4$, and (c) $2.0$.  In Fig.~\ref{fig:mudep}(a), one can clearly identify
the chiral Majorana mode, which goes across the figure from the bottom left
to the top right. In the absence of an external field, the chiral Majorana
mode has a linear dispersion in the wide range of $\mu$ (not shown) and its
dispersion is given by $E_\mathrm{CM} \sim \mu \, {\rm sgn}[M_0]$.  Namely,
the direction of the magnetization in the FI determines the direction of
the edge current\cite{TYN09}. 

%%%%%%%%%%%%%%%%%%%%%%%%%%%%%%%( mu dep )%%%%%%%%%%%%%%%%%%%%%%%%%%%%%%%%%%%%%%%
\begin{figure}[tb]
	\begin{center}
		\hspace{0.0cm}\includegraphics[width=0.48\textwidth]{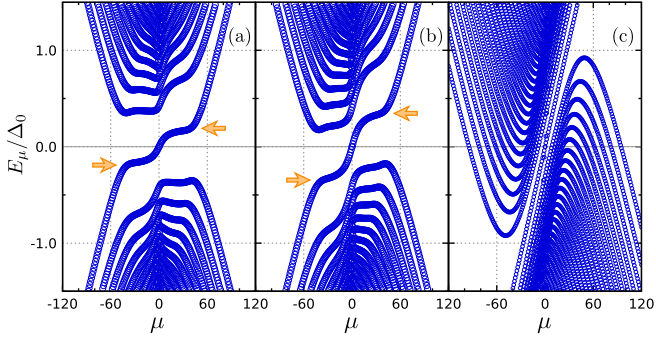}
		\caption{Angular momentum dependences of the energy spectrum for the
		chemical potential (a) $\mu_F = 0.2\Delta_0$, (b) $0.4 \Delta_0$, and
		(c) $2.0\Delta_0$. The magnetic length is set to $\ell_B = 0.08 R_c$.
		The plateaus emerge in the presence of an external field. In
		particular, the chiral Majorana bound state includes
		the $n=0$ Landau level, which is indicated with arrows in the panels
		(a) and (b). } \label{fig:mudep}
	\end{center}
\end{figure}
%%%%%%%%%%%%%%%%%%%%%%%%%%%%%%%( mu dep )%%%%%%%%%%%%%%%%%%%%%%%%%%%%%%%%%%%%%%%

%%%%%%%%%%%%%%%%%%%%%%%%%%%%%%%( LDOS20 )%%%%%%%%%%%%%%%%%%%%%%%%%%%%%%%%%%%%%%%
\begin{figure}[tb]
	\begin{center}
		\hspace{0.0cm}\includegraphics[width=0.49\textwidth]{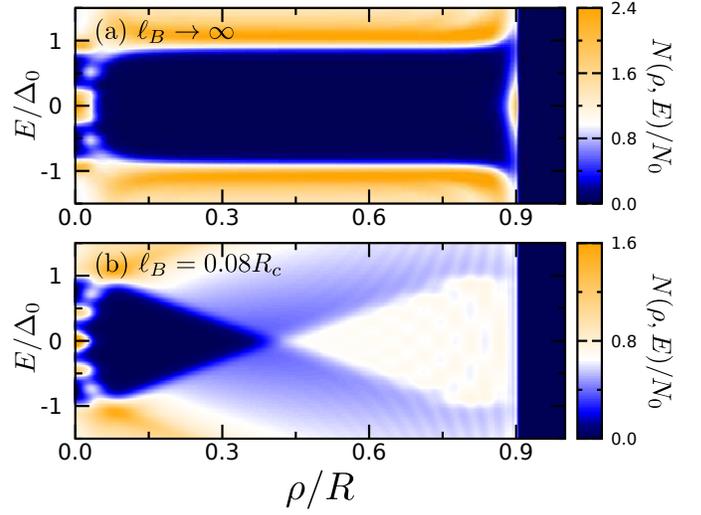}
		\caption{LDOS in the SC/TI hybrid system at $\mu_F = 2.0\Delta_0$.  The
		other parameters are set as the same values as those used in
		Fig.~\ref{fig:ldos_fe02}. {When the chemical potential $\mu_F$ is not
		sufficiently small, the exterior Majorana state is clearly seen
		only in the absence of an external magnetic field is shown in (a). Under an
		external field, the contributions from the continuum states smear out the
		peak from the exterior Majorana state as shown in (b).} }
	\label{fig:ldos_fe20}
	\end{center}
\end{figure}
%%%%%%%%%%%%%%%%%%%%%%%%%%%%%%%( LDOS20 )%%%%%%%%%%%%%%%%%%%%%%%%%%%%%%%%%%%%%%%

An external magnetic field modifies the energy spectrum through the
emergence of the Landau quantizations \cite{LL1,LL2,LL3}. In the energy
spectrum, almost-flat plateaus corresponding to the Landau levels of the
normal state emerge (indicated with arrows in Fig.~\ref{fig:mudep}). These
Landau levels are given by $E=E^{\mathrm{R}}_{\pm,n}\equiv \pm\hbar v_F
\sqrt{2n}/\ell_B - \mu_F$ and $E=-E^{\mathrm{R}}_{\pm,n}$ for particle and
hole branches, respectively, where the particle (hole) branches arise for
$\mu<0$ ($\mu>0$) when $H_z^\textrm{ext}>0$.  We note that the $\pm
E^{\mathrm{R}}_{n=0}$ plateaus emerge in a chiral Majorana mode (see
Appendix~\ref{sec:LL} for details).  However, the linear dispersion with
respect to $\mu$ remains at around $\mu=0$.  The sign of the slope around
$\mu = 0$ is determined by the direction of an applied magnetic field
$H_z^{\mathrm{ext}}$.  These dispersive states make the characteristic
V-shape peak in the $\rho$\,-$E$ space. 
% \footnote{

{
The position where the exterior MBS appears is an effective boundary
between a TSC and a topologically trivial material. Applying a strong
magnetic field (i.e., $\ell_B < R_s$), a TI becomes gapped even in the
normal state due to the Landau quantization. Introducing an $s$-wave pair
potential, Landau levels in the energy range
$|E^{\mathrm{R}}_{\pm,n}| \leq \Delta_0$ open a superconducting energy gap,
and contributes a topological superconductivity. The radius of the
topological superconducting region, in this case, depends on $\ell_B$
because $\ell_B$ determines the radius the Landau orbit and the energy of
the Landau level. 
As a results, the region with $\rho < r^*$ becomes a TSC, whereas that with
$\rho > r^*$ does not become a TSC but remains to be a normal
(non-topological) insulator.  Although there is no actual boundary such as
interfaces, there is a ``boundary'' on the radius $r^*$ at which an
exterior MBS appears. 
}
%}. 

The Landau quantization also modifies the continuum levels. For the same
magnetic field, the shift of the continuum states becomes more significant
for a larger chemical potential [Fig.~\ref{fig:mudep}(c)]. As the chemical
potential increases, the plateaus corresponding to the $n=1$ states
approach to the Fermi level.  At {large enough} chemical potential (e.g.,
$\mu_F=2.0\Delta_0$), the energy gap between the continuum states
completely disappears as shown in Fig.~\ref{fig:mudep}(c). The
contributions from the continuum states smear out the peak from the
exterior MBS in LDOS.  The LDOS for $\mu_F = 2.0\Delta_0$ are
shown in Fig.~\ref{fig:ldos_fe20}.  The two zero-energy states at the
vortex core and the SC/FI boundary can be clearly seen when $\ell_B
\rightarrow \infty$ [Fig.~\ref{fig:ldos_fe20}(a)], whereas the exterior
mode cannot be identified when $\ell_B=0.08R_c$ because the energy gap is
closed in the region of $\rho>0.4R_c$ [Fig.~\ref{fig:ldos_fe20}(b)].  We
have confirmed that the exterior MBS is located at $\rho =
0.85R_c$ in this magnetic field.  Namely, when the chemical potential is
too large, it is difficult to observe the exterior MBS through a
zero-energy peak even though its position can be controlled by an applied
field.
{
Hence, we need to set the chemical potential close to the Dirac point in
order to avoid contribution from non-zero-energy states}
\footnote{{The condition obtained here is different from that
derived in Ref.~\onlinecite{Akzyanov16}, which discusses the condition for
the energy gap due to the hybridization between vortex and exterior MBSs to
be zero or small enough.}
}. 
We also carry out the same simulations for larger systems such as $R_c =
150\xi_0$, and confirm that the above conclusion does not change
qualitatively.

%%%%%%%%%%%%%%%%%%%%%%%%%%%%%%%( sr ldos ) %%%%%%%%%%%%%%%%%%%%%%%%%%%%%%%%%%%%%%
\begin{figure}[tb]
	\begin{center}
		\hspace{0.0cm}\includegraphics[width=0.48\textwidth]{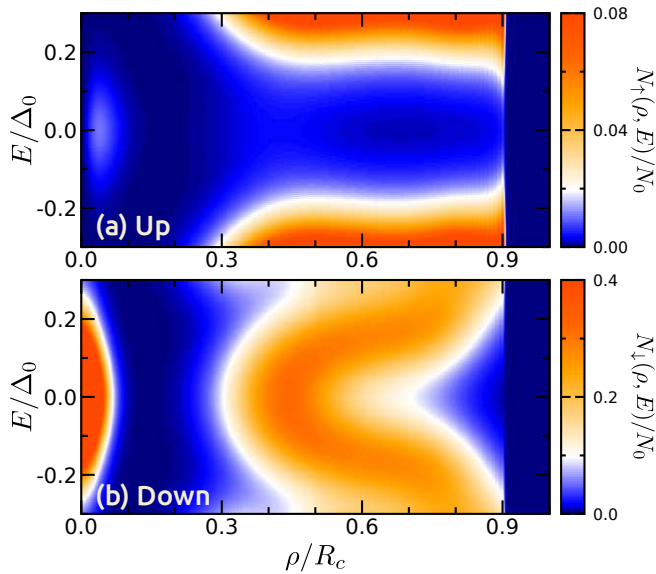}
		\caption{Spin-resolved LDOS (a) $N_\uparrow$ and (b) $N_\downarrow$ in
		the SC/TI hybrid system at $\mu_F = 0.2 \Delta_0$ and $\ell_B=0.08R_c$.
		The other parameters are set as the same values as those used in
		Fig.~\ref{fig:ldos_fe02}. {The amplitude of $N_{\downarrow}$ is
		much larger than that of $N_{\uparrow}$ at the place where the exterior
		and vortex Majorana states appear, meaning that these Majorana bound
		states are spin-polarized.} }
	\label{fig:sr_ldos}
	\end{center}
\end{figure}
%%%%%%%%%%%%%%%%%%%%%%%%%%%%%%%( sr ldos ) %%%%%%%%%%%%%%%%%%%%%%%%%%%%%%%%%%%%%%

\begin{table}[tb]
  \begin{center}
		\caption{Dependence of the majority spins of the vortex and exterior
		Majorana states on the external-field direction and the vorticity.}
    \label{tab}
    \vspace{2mm}
    \begin{tabular}{cccc} 
    \toprule
     ~$H_z^{\mathrm{ext}}$~ & ~$\ell$~ & ~Vortex MBS~ & ~Exterior MBS~ \\[-3mm] \\[0mm] 
    \hline
    \\[-2mm]
    $+$ & $+1$ & ~Down~ & ~Down~ \\[2mm] 
    $-$ & $+1$ & ~Down~ & ~Up  ~ \\[2mm]
    $+$ & $-1$ & ~Up  ~ & ~Down~ \\[2mm] 
    $-$ & $-1$ & ~Up  ~ & ~Up  ~ \\[1mm]
    \bottomrule
    \end{tabular}
  \end{center}
\end{table}

Next, we discuss the spin structure of the MBSs. The
spin-resolved LDOS for the up-spin $N_\uparrow$ and for down-spin
$N_\downarrow$ are shown in Fig.~\ref{fig:sr_ldos}, where we set $\mu_F =
0.2\Delta_0$ and $\ell_B = 0.08 R_c$.  As shown in Fig.~\ref{fig:sr_ldos},
the exterior MBS is strongly polarized.  The LDOS for the down
spin $N_\downarrow$ has a zero-energy peak around $\rho = 0.4 R_c$ and
exhibits the characteristic V-shaped peak structure in the $\rho$\,-$E$
space.  On the other hand, the amplitude of {$N_\uparrow$} at the same
place is totally small.  In the case of a TI surface, the wavefunctions for
the $n=0$ Landau level are fully spin polarized (see Appendix~\ref{sec:LL}
for details).  Correspondingly, the chiral Majorana mode which is
constituted from the $n=0$ Landau levels are spin polarized as well
\footnote{{We also calculate the spin density of
the whole system (not shown), and find that it is almost unpolarized since the
contribution from the exterior MBS is negligible compared with that from the
continuum states.  Therefore, to observe the spin-polarization of the MBS,
we need a spin- and energy-resolved measurement, which can be realised by
replacing a metallic STS tip by a ferromagnetic one. }}.

The majority spin of the exterior MBS is determined by
$\mathrm{sgn} [H_z^\mathrm{ext}]$ because the direction of the spin
polarization for the $n=0$ Landau level is determined by
$\mathrm{sgn}[H_z^\mathrm{ext}]$. Additionally, the vortex MBS
is also strongly spin polarized. As shown in Fig.~\ref{fig:sr_ldos}, the
height of the peak around $\rho \simeq 0.05 R_c$ in $N_\uparrow$ is much
smaller than the peak at $\rho=0$ in $N_\downarrow$.  The majority spin of
the vortex MBS is determined by the vorticity $\ell$. This
behavior is consistent with the previous discussions \cite{Cheng,
Rakhmanov, Akzyanov16}.  The relation among the spin polarization of
MBSs, the external magnetic field, and the vorticity is
summarized in Table~\ref{tab}. 

%%%%%%%%%%%%%%%%%%%%%%%%%%%%%%%%%%%%%%%%%%%%%%%%%%%%%%%%%%%%%%%%%%%%%%%%%%%%%%%%
%%%%%%%%%%%%%%%%%%%%%%%%%%%%%%%%%%%%%%%%%%%%%%%%%%%%%%%%%%%%%%%%%%%%%%%%%%%%%%%%
\subsubsection{Odd-frequency Cooper pairs}

%%%%%%%%%%%%%%%%%%%%%%%%%%%%%%%( Pair amp. )%%%%%%%%%%%%%%%%%%%%%%%%%%%%%%%%%%%%%%%
\begin{figure}[tb]
	\begin{center}
		\hspace{0.0cm}\includegraphics[width=0.40\textwidth]{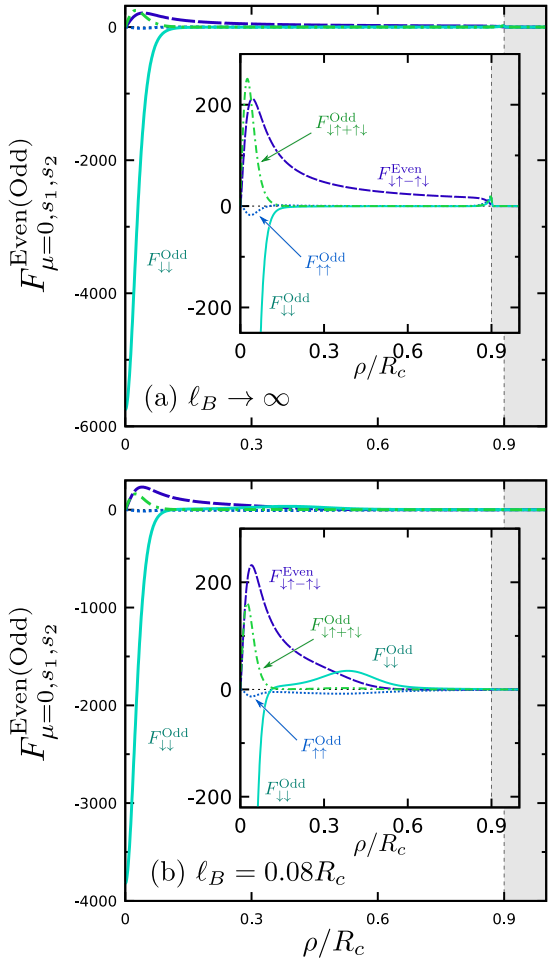}
		\caption{Pair amplitudes in the SC/TI hybrid system with (a) $\ell_B
		\rightarrow \infty$ and (b) $\ell_B = 0.08R_c$. We show the magnified
		figures in the insets. Here only the contributions from $\mu = 0$ are shown
		because the Majorana states are the eigenfunctions for $\mu=0$.  The
		chemical potential and the Matsubara frequency are set to $\mu_F =
		0.2\Delta_0$ and $\omega_n = 0.1\Delta_0$, respectively. The other
		parameters are the same as those used in Fig.~\ref{fig:ldos_fe02}.
		{ 
		When
		$H^{\mathrm{ext}}=0$, the odd-frequency Cooper pairs are localized at the
		SC/FI boundary but not spin-polarized. When $H^{\mathrm{ext}} \neq 0$, the
		odd-frequency pairs move into the superconducting region as the exterior
		Majorana state does. Simultaneously, the amplitude of
		$F^{\mathrm{Odd}}_{\downarrow \downarrow}$ is much larger than those of
		$F^{\mathrm{Odd}}_{\uparrow \uparrow}$ and $F^{\mathrm{Odd}}_{\downarrow
		\uparrow + \uparrow \downarrow}$ there. Namely, the odd-frequency pairs are
		spin-polarized, reflecting that the exterior Majorana state is spin-polarized. }}
	\label{fig:agf}
	\end{center}
\end{figure}
%%%%%%%%%%%%%%%%%%%%%%%%%%%%%%%( Pair amp. )%%%%%%%%%%%%%%%%%%%%%%%%%%%%%%%%%%%%%%%

We discuss the correspondence between the MBSs and
odd-frequency Cooper pairs. 
{
Superconducting phenomena near the Fermi level
can be interpreted by two different physical descriptions: quasiparticle
description and Cooper-pair description. Connecting these two descriptions,
one can see that, when a MBS appears, odd-frequency Cooper
pairs must appear behind it (see, for example,
Ref.~[\onlinecite{Asano2013}]). So
far, phenomena related to odd-frequency Cooper pairs accompanied by vortex
MBSs have been discussed~\cite{Yokoyama_2008_PRB, Tanuma_PRL_2009, Daino_PRB_2012}. For example, the frequency
symmetry of Cooper pairs can be
clarify by means of the local
Josephson coupling \cite{Yokoyama_2008_PRB, Kashuba_PRB_2017}.
Such results hold also for exterior MBSs. 
}

We show the $\rho$-dependences of the anomalous Green's functions in
Fig.~\ref{fig:agf}, where we fix the center of the azimuthal angle $\phi_c
=0$ and $\mu_F = 0.2\Delta_0$, and the region attached to the FI is shaded.
The magnetic length is set to (a) $\ell_B \rightarrow \infty$ and (b) $0.08
R_c$.  The imaginary (real) part of the even-frequency (odd-frequency)
components are not shown because they are negligibly small in this scale of
the plot. We show the magnified figures in the insets. 

The results for $\ell_B \rightarrow \infty$ are shown in
Fig.~\ref{fig:agf}(a).  The MBS appears at the SC/FI
boundary as shown in Fig.~\ref{fig:ldos_fe02}. Correspondingly, the
odd-frequency spin-triplet Cooper pairs appear there. Here, all of the
triplet components (i.e., 
$F^{\mathrm{Odd}}_{\uparrow \uparrow}$, 
$F^{\mathrm{Odd}}_{\downarrow \uparrow + \uparrow \downarrow},$ and 
$F^{\mathrm{Odd}}_{\downarrow \downarrow}$)
have almost the same amplitudes at the SC/FI boundary, meaning that the
synthetic spins of the odd-frequency Cooper pairs are not aligned.  At the
vortex core, there are peaks of the odd-frequency components\cite{
Yokoyama_2008_PRB, Tanuma_PRL_2009, Daino_PRB_2012} corresponding to the
vortex MBSs. In particular, $F_{\downarrow \downarrow
}^\mathrm{Odd}$ becomes nonzero even at $\rho=0$. 
{The
conventional even-frequency pair amplitude is known to be suppressed at the
place where Andreev bound states appear. } \cite{Yokoyama_2008_PRB,
Tanuma_PRL_2009, Higashitani_PRL_2013, SIS1, Daino_PRB_2012, SIS3}.
We have confirmed that the spatial profile of
$F^{\mathrm{Even}}_{\downarrow \uparrow - \uparrow \downarrow}$
for $\ell_B \rightarrow \infty$ becomes similar to
$\Delta(\rho)$ given in Eq.~\eqref{eq:Delta_rho} (i.e.,
$\Delta(\rho)$ is suppressed at around $\rho=0$ and $R_s$, by summing up
with respect to $\mu$). 

In the presence of an external magnetic field, the odd-frequency pairs at
the SC/FI boundary move inside the superconducting region. At $\ell_B =
0.08R_c$, only the $F_{\downarrow\downarrow}^{\mathrm{Odd}}$ component has
a peak around $\rho\approx 0.4R_c$. This radius is exactly where the
spin-polarized exterior MBS has an large amplitude.  The
direction of spin polarization of the odd-frequency pairs is determined by
the spin polarization of the $n=0$ ``relativistic'' Landau level\cite{LL3}
(i.e., determined by $\textrm{sgn}[H_z^\textrm{ext}]$).
{
The place where the odd-frequency pair amplitude emerges is actually a
boundary between a superconducting region and an insulating region due to
the Landau quantization: Even though the superconducting pairing is
proximity induced at $\rho<R_s$ from an external SC, the even-frequency
pair amplitude is highly suppressed outside the region surrounded by the
peak of the odd-frequency pair amplitude as shown in Fig.~\ref{fig:agf}(b). 
}

%%%%%%%%%%%%%%%%%%%%%%%%%%%%%%%%%%%%%%%%%%%%%%%%%%%%%%%%%%%%%%%%%%%%%%%%%%%%%%%%
%%%%%%%%%%%%%%%%%%%%%%%%%%%%%%%%%%%%%%%%%%%%%%%%%%%%%%%%%%%%%%%%%%%%%%%%%%%%%%%%
\subsection{Rashba superconductor}
\begin{figure}[tb]
	\begin{center}
		\hspace{0.0cm}\includegraphics[width=0.48\textwidth]{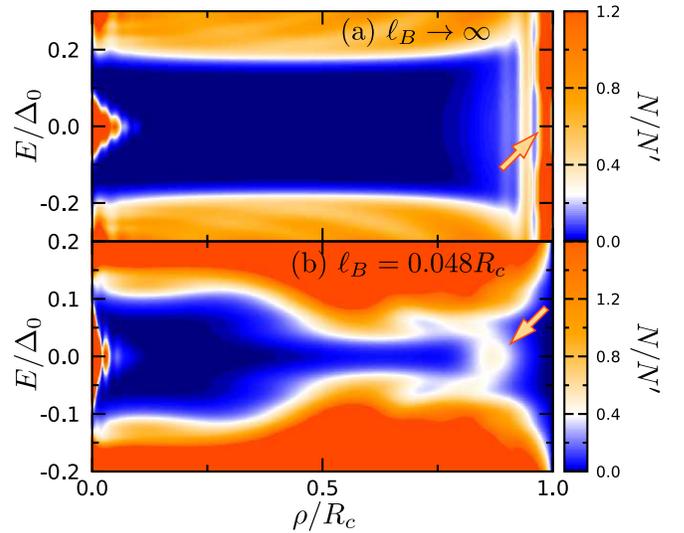}
		\caption{LDOS in the Rashba SC at (a) $\ell_B \rightarrow \infty$ and
		(b) $\ell_B = 0.048R_c$.  The edge of the SC is located at $\rho =
		R_c$. The results are normalized by $N'$, which is the LDOS at $E =
		0.4\Delta_0$ and $\rho = 0.2R_c$ in the superconducting state without
		an external field. The parameters are set to $\mu_F = 0.2\mu_F$, $R_c =
		180\xi_0$, $\ell=1$, $M_0 = 1.2\Delta_0$, and $m_0 \lambda^2 =
		\Delta_0$.
		{
		The exterior Majorana state is located at the
		edge of the system in (a), whereas it moves to $\rho \approx 0.85R_c$
		in (b). The arrows indicate the position of the exterior Majorana state.} 
		}
	\label{fig:RM_zeldos1}
	\end{center}
\end{figure}

The Rashba SCs can also host the MBSs at the vortex core
and at the edge as in the case of the SC/TI hybrid system.  We discuss the
numerical results for the Rashba SC satisfying the topological criterion
\cite{STF09, oreg10, lutchyn10, SAU, Yamakage} $M_0^2 > \mu_F^2 +
\Delta_0^2$. We show the LDOS for $\ell_B \rightarrow \infty$ and $\ell_B =
0.048 R_c$ in Figs.~\ref{fig:RM_zeldos1}(a) and \ref{fig:RM_zeldos1}(b),
respectively, where the parameters are set to $\mu_F = 0.2 \Delta_0$, $R_c
= 180 \xi_0$, $m_0 \lambda^2 = \Delta_0$, and $M_0 = 1.2\Delta_0$.

%%%%%%%%%%%%%%%%%%%%%%%%%%%%%%%( Figure LDOS )%%%%%%%%%%%%%%%%%%%%%%%%%%%%%%%%%%%%%%%

\begin{figure}[tb]
	\begin{center}
		\hspace{0.0cm}\includegraphics[width=0.48\textwidth]{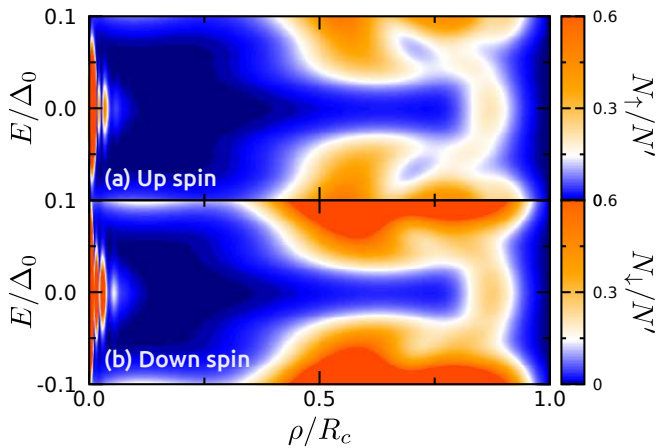}
		\caption{Spin-resolved LDOS (a) $N_\uparrow$ and (b) $N_\downarrow$ in
		the Rashba SC at $\ell_B=0.048R_c$.  The edge of the SC is located at
		$\rho = R_c$. The results are normalized by $N'$ defined in
		Fig.~\ref{fig:RM_zeldos1}.  The other parameters are the same as those
		used in Fig.~\ref{fig:RM_zeldos1}. 
		{
		Contrary to the case in
		Fig.~\ref{fig:sr_ldos}, the exterior Majorana state at $\rho = 0.85R_c$
		consists of the up- and down-spin quasiparticles. }
		}
	\label{fig:RM_srldos}
	\end{center}
\end{figure}

\begin{figure}[tb]
	\begin{center}
		\hspace{0.0cm}\includegraphics[width=0.44\textwidth]{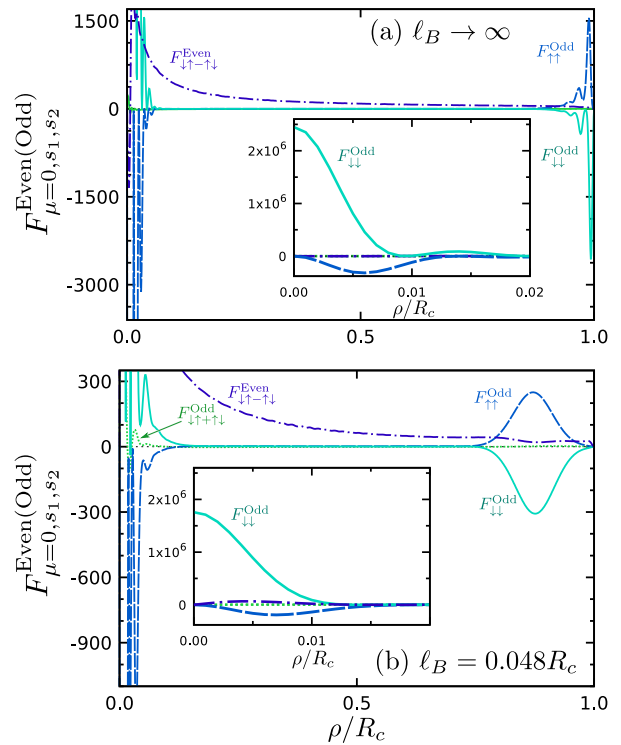}
		\caption{Pair amplitudes in the Rashba SC at $\mu = 0$ with (a) $\ell_B
			\rightarrow \infty$ and (b) $\ell_B = 0.048R_c$. Here the results are
			shown with limited vertical axes to focus on the exterior Majorana
			state. The pair amplitudes near the vortex are shown in the insets.
			The Matsubara frequency are set to $\omega_n = 0.01\Delta_0$ which is
			smaller than the effective gap. The other parameters are the same as
			those used in Fig.~\ref{fig:RM_zeldos1}. 
			{
			Reflecting that the exterior Majorana state is not spin-polarized in
			a Rashba SC, the odd-frequency Cooper pairs are not spin-polarized 
		(i.e., $F_{\uparrow \uparrow}^{\mathrm{Odd}}$ and $F_{\downarrow
\downarrow}^{\mathrm{Odd}}$ have almost the same amplitude).  }
			}
	\label{fig:RM_agf}
	\end{center}
\end{figure}
%%%%%%%%%%%%%%%%%%%%%%%%%%%%%%%( Schematic Figure )%%%%%%%%%%%%%%%%%%%%%%%%

In the absence of an external field, the MBSs are located
at the center of the core and at the edge of a system. These Majorana
states appear as zero-energy peaks in LDOS at $\rho = 0$ and $\rho = R_c$.
Here, the magnitude of the energy gap is smaller than $\Delta_0$ because
this energy gap is caused by the effective triplet pairings
\cite{SAU,Yamakage}. The result in the presence of a magnetic field is
shown in Fig.~\ref{fig:RM_zeldos1}(b). The subgap states do not appear at
the edge of the system but at $\rho \approx 0.85R_c$.  Namely, by applying
an external field, the exterior MBS for the Rashba SC moves
inside a superconducting region as seen in an SC/TI hybrid system. We
therefore conclude that the controllability of the exterior MBS
is not restricted to materials with linear dispersion but a general
property of two-dimensional TSCs. 

However, there are two qualitative differences between exterior MBS in a Rashba SC and that in a SC/TI hybrid system.  First, the
exterior MBS is not spin-polarized as shown in
Fig.~\ref{fig:RM_srldos}, where the spin-resolved LDOS in a Rashba SC are
plotted. Reflecting the quadratic dispersion, the wavefunction of the $n=0$
Landau level is not spin-polarized in a Rashba SC (see
Appendix~\ref{sec:LL}). As a result, the exterior MBS and {the}
corresponding triplet odd-frequency Cooper pairs are not spin-polarized.
The pair amplitudes are shown in Fig.~\ref{fig:RM_agf}. In the absence of
an external magnetic field, the pair amplitudes have {peaks} at the edge of
the system. When $\ell_B = 0.048R_c$, on the other hand, {the peaks} move
inside {the system as seen in the case of} SC/TI hybrid systems.  However,
$F_{\uparrow \uparrow}^{\mathrm{Odd}}$ and $F_{\downarrow
\downarrow}^{\mathrm{Odd}}$ have similar spatial profiles though their
signs are opposite.  In other words, the odd-frequency Cooper pairs are not
spin polarized in a Rashba SC. 
 
Second, the characteristic V-shape peak in the LDOS cannot be identified in
Fig.~\ref{fig:RM_zeldos1}(b).  In a Rashba SC, a energy gap is smaller
compared with that in an SC/TI hybrid system.  As a result, there is only a
few subgap ``edge'' states.  Although one can increase the number of subgap
states by increasing a system size, in that case, the continuum states
approach to the Fermi level in the presence of an external magnetic field,
and smear out the exterior MBS. Accordingly in a Rashba SC, fine
tunings of many parameters (e.g., radius of a Rahba SC and an external
magnetic field) are required to observe the exterior MBS.

\section{Conclusion}
\label{sec:conclusion}

We have theoretically studied the quasiparticle spectrum of two-dimensional
topological superconductors hosting a vortex under an external magnetic
field.  We have mainly considered the surface of a topological insulator to
which a superconductor with a superconducting vortex is proximity-coupled.
We have obtained the spin-resolved local density of states by solving the
Bogoliubov-de Gennes equation, and shown that the exterior Majorana state
can be observed as a peak in the local density of states, which form a
characteristic V-shaped peak in radius-energy space. Carrying out the same
simulations for a Rashba superconductor, we have concluded that the shift
in the real space of the exterior Majorana state by applying a magnetic
field is general property of two-dimensional topological superconductors. 

Moreover, we have elucidated {that} there are qualitative difference
between the exterior Majorana state in a topological-insulator surface and
that in a Rashba superconductor.  In the former case, an exterior Majorana
state is fully spin polarized reflecting the spin-polarization of the $n=0$
relativistic Landau level. On the other hand, in a Rashba superconductor,
an exterior Majorana state is not spin polarized because of a conventional
quadratic dispersion relation. 

This difference affects on the spin structure of induced spin-triplet
odd-frequency $s$-wave Cooper pairs. On a topological-insulator surface,
corresponding to the spin polarization of exterior Majorana state, the
odd-frequency Cooper pairs are fully spin polarized as well. For a Rashba
superconductor, on the other hand, the spin of odd-frequency pairs is not
polarized.  

We have also shown that energy dependence of the local density of states
around the exterior Majorana state strongly depends on the chemical
potential. When the chemical potential is not sufficiently small, continuum
states come close to the Fermi level because of the energy shift due to an
external magnetic field. As a result, fine-tuning of the chemical potential
is necessary to experimentally observe the position shift of exterior
Majorana states. 

\begin{acknowledgments}
The authors would like to thank S.~Kashiwaya and S.~Tamura for useful
discussions.  This work was supported by JST-CREST (Grant No.~JPMJCR16F2),
and JSPS KAKENHI Grants No.~JP16H00989 and No.~JP15K17726. 
Y.~T. is supported by a Grant-in-Aid for Scientific Research on Innovative
Areas ``Topological Materials Science'' (KAKENHI Grant No.~JP15H05853) from
JSPS of Japan, a Grant-in-Aid for Scientific Research B (Grant
No.~JP15H03686), a Grant-in-Aid for Challenging Exploratory Research (Grant
No.~JP15K13498) from the Ministry of Education, Culture, Sports, Science,
and Technology, Japan  (MEXT); Japan-RFBR JSPS Bilateral Joint Research
Projects/Seminars.
\end{acknowledgments}

\appendix

\section{Hamiltonian for a superconductor/topological insulator hybrid
system in terms of the Bessel functions}
\label{sec:app_A}

In rotational-symmetric systems, the solutions can be well described in
terms of the Bessel functions \cite{Cheng,TKawakami}.  We thus introduce
the Bessel functions as the basis for the real space as 
\begin{align}
	\Psi_\mu(\rho) = 
	\sum_{j=1}^{j_{\textrm{max}}}
	\left( \begin{array}{c}
		\psi        _{  \uparrow, \mu, j} \phi_{\mu_{--},j}\left( \alpha_{\mu_{--},j}{\rho}/{R_c}\right)\\[1.4mm] 
		\psi        _{\downarrow, \mu, j} \phi_{\mu_{-+},j}\left( \alpha_{\mu_{-+},j}{\rho}/{R_c}\right)\\[1.4mm]
		\psi^\dagger_{  \uparrow, \mu, j} \phi_{\mu_{+-},j}\left( \alpha_{\mu_{+-},j}{\rho}/{R_c}\right)\\[1.4mm]
		\psi^\dagger_{\downarrow, \mu, j} \phi_{\mu_{++},j}\left( \alpha_{\mu_{++},j}{\rho}/{R_c}\right)\\
	\end{array}	\right), 
\end{align}
where we define $\mu_{s_3 s_4}=( 2\mu + s_3 \ell +s_4 1)/2,\, (s_3, s_4 =
\pm)$ and
\begin{align}
	\phi_{\mu,j} = 
	\frac{\sqrt{2}}{ R_c J_{\mu+1}   \left( \alpha_{\mu,j} \right) }
	J_{\mu}   \left( \alpha_{\mu  ,j} \cfrac{\rho}{R_c} \right), 
\end{align}
with $J_{\mu}(\rho)$ being the Bessel functions with the order $\mu$, and
$\alpha_{\mu  ,j} $ is the $j$-th zero of $J_{\mu}$.  These functions
$\phi_{\mu,j}$ satisfy 
\begin{align}
  \int_0^{R_c}  
  \phi_{\mu j} \left( \alpha_{\mu,j } \cfrac{\rho}{R_c} \right) 
  \phi_{\mu j'} \left( \alpha_{\mu,j'} \cfrac{\rho}{R_c} \right) 
	\rho d\rho = \delta_{jj'}, 
\end{align}
where we have used 
$
\int_0^{R_c} \left[ 
J_{\mu} \left( \alpha_{\mu  ,j} {\rho}/{R_c} \right) 
\right]^2 \rho d\rho = 
\left[ R_c J_{\mu+1} \left( \alpha_{\mu  ,j} \right) \right]^2 / 2
$. 

In the numerical calculations, we introduce the cutoff in the summation of
$j$. The maximum value is denoted by $j_{\mathrm{max}}$, and it is set to
$j_{\mathrm{max}} = 200 $ for the TI surface.  We assume the magnetic field
is spatially homogeneous $H_z(\rho) = (\boldsymbol{\nabla} \times
\boldsymbol{A})_z = H_z^{\mathrm{ext}}$.  This magnetic field can be
described by the vector potential $ A_\phi = H_z^{\mathrm{ext}} \rho /2 $
($H_z = [ \partial_\rho (\rho A_\phi)]/\rho $).  We adopt the vector
potential $ \tilde{A}_\phi = \mathcal{S} {\rho}/({2 \ell_B^2}) $ where
$\ell_B = (\hbar c / |e H_z^{\mathrm{ext}}|)^{1/2}$ is the magnetic length
and $\mathcal{S} = \mathrm{sgn} [H_z^{\mathrm{ext}}]$. With this basis, the
Hamiltonian becomes
\begin{widetext}
\begin{align}
	\mathcal{H}& =
		\sum_{\mu } \sum_{i,j} 
		\begin{pmatrix}
      \psi_{  \uparrow,\mu,i}^\dagger & 
      \psi_{\downarrow,\mu,i}^\dagger & 
		  \psi_{  \uparrow,\mu,i}         & 
      \psi_{\downarrow,\mu,i} 
    \end{pmatrix} 
		\check{U}_1^\dagger 
		\notag \\
		&\times
	  \left[ \begin{array}{cccc}
			M^{(\mu_{--})}_{ij} -\mu_F \delta_{i,j} &
			\hspace{-3mm}
			\hbar v_F \Big( K^{(-)}_{\mu_{-+},i,j} + A^{(-)}_{\mu_{-+},i,j} \Big) &
			0 &
			{\Delta}_{\mu_{--},\mu_{+-},i,j} \\[3mm]
			\hspace{-1mm}
			\hbar v_F \Big( K^{(+)}_{\mu_{--},i,j} + A^{(+)}_{\mu_{--},i,j} \Big) &
			-M^{(\mu_{-+})}_{ij} -\mu_F \delta_{i,j} &
			-{\Delta}_{\mu_{-+},\mu_{++},i,j} &
			0 \\[3mm]
			 0 &
			-{\Delta}_{\mu_{++},\mu_{-+},i,j} &
			-M^{(\mu_{++})}_{ij} + \mu_F \delta_{i,j} &
			%\left( \partial_\rho  +\cfrac{\mu+1}{\rho} \right) \\
			\hspace{-3mm}
			\hbar v_F \Big( K^{(+)}_{\mu_{+-},i,j} - A^{(+)}_{\mu_{+-},i,j} \Big) \\[3mm]
			{\Delta}_{\mu_{+-},\mu_{--},i,j} &
			0 &
			\hspace{-3mm}
			\hbar v_F \Big( K^{(-)}_{\mu_{++},i,j} - A^{(-)}_{\mu_{++},i,j} \Big) &
			M^{(\mu_{+-})}_{ij} + \mu_F \delta_{i,j}  \\
	 \end{array} \right] \notag \\
	 & \times \check{U}_1 
   \begin{pmatrix} 
      \psi_{  \uparrow,\mu,j} & 
      \psi_{\downarrow,\mu,j} & 
		  \psi_{  \uparrow,\mu,j}^\dagger & 
      \psi_{\downarrow,\mu,j}^\dagger 
   \end{pmatrix}^{\rm T}, %\\[4mm]
\end{align}
where $\check{U}_1=\text{diag} \left[ \,e^{i\pi/4}~~e^{-i\pi/4}~
~e^{-i\pi/4}~ ~e^{i\pi/4}\, \right]$.  The each term is described as 
\begin{align}
	K^{(+)}_{\mu,i,j} &= \int_0^{R_c}
	\frac{\alpha_{\mu,j}}{R_c}
	\frac{2}{R_c^2}
	\frac
	{J_{\mu+1}\left( \alpha_{\mu+1,i}{\rho}/{R_c} \right)}
	{J_{\mu+2}\left( \alpha_{\mu+1,i} \right)}
	\frac
	{J_{\mu+1}\left( \alpha_{\mu,j}{\rho}/{R_c} \right)}
	{J_{\mu+1}\left( \alpha_{\mu,j} \right)}
	\rho d\rho, \\[2mm]
	K^{(-)}_{\mu,i,j} &= \int_0^{R_c}
	\frac{\alpha_{\mu,j}}{R_c}
	\frac{2}{R_c^2}
	\frac
	{J_{\mu-1}\left( \alpha_{\mu-1,i}{\rho}/{R_c} \right)}
	{J_{\mu  }\left( \alpha_{\mu-1,i} \right)}
	\frac
	{J_{\mu-1}\left( \alpha_{\mu}{\rho}/{R_c} \right)}
	{J_{\mu+1}\left( \alpha_{\mu} \right)}
	\rho d\rho, %\\[2mm]
\end{align}
%\end{widetext}
\begin{align}
	A^{(+)}_{\mu,i,j} &= \int_0^{R_c}
	\frac{\mathcal{S}}{2 \ell_B^2}
	{\phi_{\mu+1}\left( \alpha_{\mu+1,i}{\rho}/{R_c} \right)}
	{\phi_{\mu  }\left( \alpha_{\mu,j}{\rho}/{R_c} \right)}
	\rho^2 d\rho, \\[1mm]
	A^{(-)}_{\mu,i,j} &= \int_0^{R_c}
	\frac{\mathcal{S}}{2 \ell_B^2}
	{\phi_{\mu-1}\left( \alpha_{\mu-1,i}{\rho}/{R_c} \right)}
	{\phi_{\mu}\left( \alpha_{\mu,j}{\rho}/{R_c} \right)}
	\rho^2 d\rho, %\\[1mm]
\end{align}
\begin{align}
	\Delta_{u,v,i,j} &= \int_0^{R_c} \Delta {(\rho)}
	{\phi_{u}\left( \alpha_{u,i}\,{\rho}/{R_c} \right)}
	{\phi_{v}\left( \alpha_{v,j}\,{\rho}/{R_c} \right)}
	\rho d\rho, \\[1mm]
  M^{(\mu)}_{ij} &= \int_0^{R_c}M {(\rho)}
	{\phi_{\mu}\left( \alpha_{\mu,i}\,{\rho}/{R_c} \right)}
	{\phi_{\mu}\left( \alpha_{\mu,j}\,{\rho}/{R_c} \right)}
	\rho d\rho. 
	\label{eq:Mag_Bes_term}
\end{align}
We diagonalize numerically this $4j_{\mathrm{max}} \times
4j_{\mathrm{max}}$ Hamiltonian for each $\mu$, and obtain the eigenfunction
$\Phi_{\mu,\nu}$ and the energy eigenvalue $E_{\mu,\nu}$. 

\section{Hamiltonian for a Rashba SC in terms of the Bessel functions}
\label{sec:app_B}
%\begin{widetext}
Introducing the Bessel functions, the Hamiltonian for the Rashba SC becomes 
\begin{align}
	\mathcal{H}& =
		\sum_{\mu } \sum_{i,j} 
		\big[\,
      \psi_{  \uparrow,\mu,i}^\dagger ~ 
      \psi_{\downarrow,\mu,i}^\dagger ~ 
      \psi_{  \uparrow,\mu,i}         ~ 
      \psi_{\downarrow,\mu,i} 
		\, \big]
		\notag \\
		&\times
	  \left[ \begin{array}{cccc}
			C^{(-+)}_{\mu_{--},i,j} +D_{\mu_{--},i,j} &
			\hspace{-1mm}
			\lambda \Big( K^{(-)}_{\mu_{-+},i,j} + A^{(-)}_{\mu_{-+},i,j} \Big) &
			0 &
			{\Delta}_{\mu_{--},\mu_{+-},i,j} \\[3mm]
			\hspace{-1mm}
			\lambda \Big( K^{(+)}_{\mu_{--},i,j} + A^{(+)}_{\mu_{--},i,j} \Big) &
			C^{(--)}_{\mu_{-+},i,j} +D_{\mu_{-+},i,j} & 
			-{\Delta}_{\mu_{-+},\mu_{++},i,j} &
			0 \\[3mm]
			 0 &
			-{\Delta}_{\mu_{++},\mu_{-+},i,j} &
			-C^{(++)}_{\mu_{++},i,j} -D_{\mu_{++},i,j} & 
			%\left( \partial_\rho  +\cfrac{\mu+1}{\rho} \right) \\
			\hspace{-1mm}
			\lambda \Big( K^{(+)}_{\mu_{+-},i,j} - A^{(+)}_{\mu_{+-},i,j} \Big) \\[3mm]
			{\Delta}_{\mu_{+-},\mu_{--},i,j} &
			0 &
			\hspace{-1mm}
			\lambda \Big( K^{(-)}_{\mu_{++},i,j} - A^{(-)}_{\mu_{++},i,j} \Big) &
			-C^{(+-)}_{\mu_{+-},i,j} -D_{\mu_{+-},i,j} \\
	 \end{array} \right] \notag \\
	&\times \left[\,
	\psi        _{\uparrow  , \mu, j} ~%\\[2mm] 
	\psi        _{\downarrow, \mu, j} ~%\\[2mm]
	\psi^\dagger_{\uparrow  , \mu, j} ~%\\[2mm]
	\psi^\dagger_{\downarrow, \mu, j} %~\\
  \, \right]^{\mathrm{T}}, 
\end{align}
where
\begin{align}
	C_{\mu,i,j}^{( \pm \pm )} &= 
	\varepsilon_{\mu,i,j}^\pm 
	%\pm M      _{\mu,i,j}, 
	\pm M_0      \delta_{i,j}, 
	\\[1mm]
	\varepsilon_{\mu,i,j}^\pm 
	&= 
	\left[ \cfrac{\hbar^2}{2m} 
		\Big( \cfrac{\alpha_{\mu,i}^2}{ R_c^2 } \pm \cfrac{\mu }{2 \ell_B^2} \Big) - \mu_F
	\right] \delta_{ij}, 	\\%
	D_{\mu,i,j} &= \int_0^{R_c}
	\frac{\hbar^2}{2 m}
	\frac{\rho^3}{4 \ell_B^4}
	{\phi_{\mu}\left( \alpha_{\mu,i}{\rho}/{R_c} \right)}
	{\phi_{\mu}\left( \alpha_{\mu,j}{\rho}/{R_c} \right)}
	d\rho.
\end{align}
\end{widetext}
The first sign in the superscript of $C_{\mu,i,j} ^{(\pm\pm)}$ corresponds
to the sign in the superscript of $\varepsilon$, and the second one does to
the sign in front of $M_0$. 

In the numerical calculation for the Rashba SC, $j_{\mathrm{max}}$ is set
to $j_{\mathrm{max}} = 200$.  Diagonalizing the Hamiltonian for each $\mu$,
we obtain the eigenfunction $\Phi_{\mu,\nu}$ and the energy eigenvalue
$E_{\mu,\nu}$. 

\section{Landau levels}
\label{sec:LL}

\subsection{Non-relativistic particle}
A non-relativistic particle has a quadratic dispersion relation. The
Hamiltonian is given by
\begin{align}
	\hat{h}_\mathrm{NR} 
	&= \left[ \frac{1}{2m_0} \left( \tilde{p}_x^2 +\tilde{p}_y^2\right)  
  -\mu_F \right] \hat{\sigma}_0, 
\end{align}
where $\tilde{\boldsymbol{p}} = \boldsymbol{p} - e \boldsymbol{A}/c $ and
the basis is taken as
$
	 \boldsymbol{\psi}             (\boldsymbol{r}) = 
	[~\psi_{\uparrow}   (\boldsymbol{r}) ~ ~
	  \psi_{\downarrow} (\boldsymbol{r}) ~]^{\mathrm{T}}
$. 
It is convenient to introduce the ladder operators $a$ and $a^\dagger$
given by
\begin{align}
	a         = \frac{\ell_B}{\sqrt{2} \hbar} ( \tilde{p}_x -i\mathcal{S}\tilde{p}_y),\hspace{6mm}%\\[2mm]
  a^\dagger = \frac{\ell_B}{\sqrt{2} \hbar} ( \tilde{p}_x +i\mathcal{S}\tilde{p}_y), 
	\label{eq:op-a}
\end{align}
where $\mathcal{S}=\textrm{sgn}[H_z^\textrm{ext}]$ and the ladder operators
satisfy $[ a , a^\dagger]_- = 1$. With these operators, the Hamiltonian
reduces to 
\begin{align}
	\hat{h}_\mathrm{NR} 
	&= \left[ \hbar \omega_c \left( a^\dagger a + \frac{1}{2} \right) -\mu_F \right]
	\hat{\sigma}_0, 
\end{align}
where $\omega_c = |e H^{\mathrm{ext}}_z| / mc$, and the energy eigenvalue
are given by
\begin{align}
	E^{\mathrm{NR}}_{n} = \hbar \omega_c \left( n + \frac{1}{2} \right) -\mu_F, 
\end{align}
with $n \ge 0$ being an integer. The corresponding eigenstates are doubly
degenerate and given by
\begin{align}
  \left[ \begin{array}{c}
    |n\rangle \\ 
    0 
  \end{array} \right], 
  \hspace{6mm}
  \left[ \begin{array}{c}
    0 \\
    |n\rangle \\ 
  \end{array} \right], 
\end{align}
where $|n \rangle$ is a number eigenstate satisfying the relations 
$a^\dagger a |n \rangle = n |n \rangle$, 
$ \langle n|n' \rangle = \delta_{nn'}$, 
$a^\dagger |n \rangle = \sqrt{n+1}|n+1\rangle$, 
$a         |n \rangle = \sqrt{n  }|n-1\rangle$, 
and $a |0 \rangle = 0$.

\subsection{Relativistic particle}
A relativistic particle has a linear dispersion relation with the
Hamiltonian given by
\begin{align}
	\hat{h}_\mathrm{R}
	= v_F \hat{\boldsymbol{\sigma}} \cdot \tilde{\boldsymbol{p}} -\mu_F  %\\	
	= \hat{h}_\mathrm{R}' -\mu_F, 
\end{align}
where $\hat{h}_\mathrm{R}' = v_F \hat{\boldsymbol{\sigma}} \cdot
\tilde{\boldsymbol{p}}$.  The $2\times 2$ matrix form of
$\hat{h}_\mathrm{R}'$ for $\mathcal{S} = +1$ is given by
\begin{align}
  \hat{h}_\mathrm{R}'
	=  
	v_F \left[ \begin{array}{cc}
			0 & 
			\tilde{p}_x - i\tilde{p}_y \\[1mm]
			\tilde{p}_x + i\tilde{p}_y & 
			0 \\
	\end{array} \right] %\\[2mm] & 
	= \frac{ \sqrt{2} \hbar v_F}{\ell_B}
	\left[ \begin{array}{cc}
			~0~ & 
			~a~ \\[1mm]
			~a^\dagger~ & 
      ~0~ \\
	\end{array} \right],
\end{align}
where $a$ and $a^\dagger$ are defined in Eq.~\eqref{eq:op-a}. Using the
fact that $(\hat{h}_\mathrm{R}')^2$ is diagonal:
\begin{align}
	[\hat{h}_\mathrm{R}']^2
	%= 2 \left( \frac{ \hbar2 v_F^2}{\ell_B^2}
	= 2 \left( \frac{ \hbar v_F}{\ell_B} \right)^2
	\left[ \begin{array}{cc}
			a^\dagger a +1& 0
			\\
			0 & 
			a^\dagger a \\
	\end{array} \right], 
\end{align}
the eigenvalues of $\hat{h}_\mathrm{R}$ are given by
\begin{align}
	E^{\mathrm{R}}_{\pm,n} = \pm \frac{ \hbar v_F}{\ell_B} \sqrt{2n} - \mu_F,
\end{align}
where $n$ is a non-negative integer. 

The eigenfunctions associated with $E^{\mathrm{R}}_{\pm,n \neq 0}$ are
given by
\begin{align}
	&\left[ \begin{array}{c}
		B_{\uparrow   ,\pm, n} \\[1mm]
	  B_{\downarrow ,\pm, n} \\
  \end{array} \right]
	= \frac{1}{\sqrt{2}}%A_n
  \left[ \begin{array}{c}
		    |n-1 \rangle \\[1mm]
	  \pm |n   \rangle \\
  \end{array} \right], 
\end{align}
and that with $E^{\mathrm{R}}_{n = 0}$ is 
\begin{align}
	&\left[ \begin{array}{c}
		B_{\uparrow  ,n=0} \\[1mm]
	  B_{\downarrow,n=0} \\
  \end{array} \right]
	= %A_0
  \left[ \begin{array}{c}
		  0         \\[1mm]
	  | 0 \rangle \\
  \end{array} \right]. 
\end{align}
Note that the $n=0$ state is fully spin-polarized. 

Having done the same calculation for $\mathcal{S} = -1$, we can obtain the
eigenfunction with $E^{\mathrm{R}}_{\pm,n \neq 0}$ are given by
\begin{align}
	&\left[ \begin{array}{c}
		C_{\uparrow   ,\pm, n} \\[1mm]
	  C_{\downarrow ,\pm, n} \\
  \end{array} \right]
	= \frac{1}{\sqrt{2}}%A_n
  \left[ \begin{array}{c}
	  \pm |n   \rangle \\[1mm]
		    |n-1 \rangle \\%[1mm]
  \end{array} \right], 
\end{align}
and that with $E^{\mathrm{R}}_{n = 0}$ is 
\begin{align}
	&\left[ \begin{array}{c}
		C_{\uparrow  ,n=0} \\[1mm]
	  C_{\downarrow,n=0} \\
  \end{array} \right]
	= %A_0
  \left[ \begin{array}{c}
	  | 0 \rangle \\[1mm]
		  0         \\
  \end{array} \right]. 
\end{align}
Therefore, we can see that the direction of the polarized spin for $n=0$
state is determined by $ \mathcal{S}$.

%%%%%%%%%%%%%%%%%%%%%%%%%%%%%%%( Schematic LL )%%%%%%%%%%%%%%%%%%%%%%%%%%%
\begin{figure}[tb]
	\begin{center}
		\vspace{8mm}\includegraphics[width=0.48\textwidth]{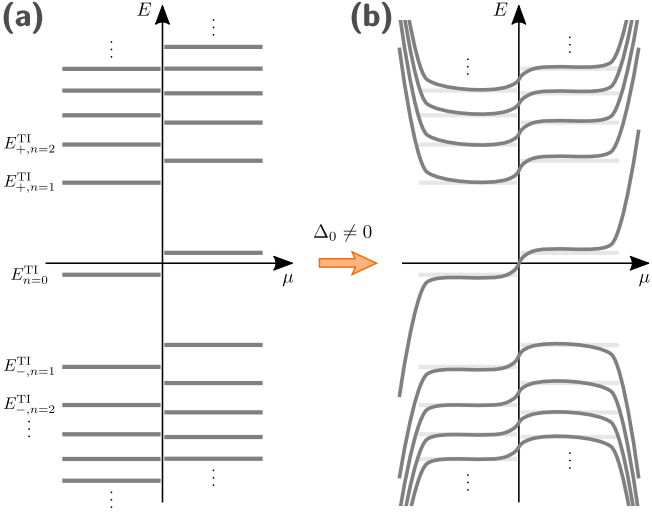}
		\caption{Schematics of Landau levels for relativistic particles. (a)
			Landau levels in the normal state ($\Delta_0=0$) for
			$\mathcal{S}=\textrm{sgn}[H_z^\textrm{ext}]=+1$, where the particle
			(hole) branches $E=E_{\pm,n}^\textrm{R} (-E_{\pm,n}^\textrm{R})$
			arise in the negative (positive) angular momentum $\mu$ region.  (b)
			In the presence of superconducting pairing ($\Delta_0\neq 0$), the
			Landau levels in (a) are smoothly connected.  The effect of the
			finite system size is also taken into account in (b).}
	\label{fig:LL-sche}
	\end{center}
\end{figure}
%%%%%%%%%%%%%%%%%%%%%%%%%%%%%%%( Schematic Figure )%%%%%%%%%%%%%%%%%%%%%

\subsection{Angular momentum}
In general, the Landau levels are degenerate with respect to the center
coordinate of the cyclotron motion. The center of the orbit ($X, Y$) is
given by 
\begin{align}
	X = x - \mathcal{S} \frac{\ell_B^2}{\hbar} \tilde{p}_y, \hspace{04mm}%\\[2mm]
  Y = y + \mathcal{S} \frac{\ell_B^2}{\hbar} \tilde{p}_x, 
\end{align}
By using $X$ and $Y$, we can introduce another ladder operators as 
\begin{align}
	 b         = \frac{1}{\sqrt{2} \ell_B} ( X + i\mathcal{S}Y ),\hspace{04mm}%\\[2mm]
   b^\dagger = \frac{1}{\sqrt{2} \ell_B} ( X - i\mathcal{S}Y ), 
\end{align}
which satisfy $[\,b, b^\dagger\,]_- = 1$.  The quantum states in the same
Landau level are specified in terms of the eigenvalue of $b^\dagger b$ as 
\begin{align}
  b^\dagger b | n, m \rangle = m | n, m \rangle, 
\end{align}
where $m$ is a non-negative integer.  The angular momentum operator is
expressed by using $b$ and $b^\dagger$ as 
\begin{align}
	L_z 
	&= (\boldsymbol{r} \times \boldsymbol{p})_z, \\[1mm]
	&= 
	\mathcal{S} \left[ ~
	   \frac{\ell_B^2}{2 \hbar} \, (\tilde{p}_x^2 + \tilde{p}_y^2)
	   -\frac{\hbar}{2 \ell_B^2} (X^2 + Y^2)~
	  \right],	\\[1mm]
	&= 
	\mathcal{S} \hbar (\,
    a^\dagger a - b^\dagger b
	  \,). 
\end{align}
That means a quantum state $| n, m \rangle$ has the angular momentum $ L_z
= \mathcal{S} \hbar ( n-m )  $. Since $n$ and $m$ are non-negative integer,
the angular momentum of the $|n=0,m \rangle$ states is restricted in $L_z <
0$ for $\mathcal{S} = +1$ and $L_z>0$ for $\mathcal{S}=-1$. 

\subsection{Bogoliubov-de Gennes formalism}
In the BdG formalism, the single-hole Hamiltonian
$-\hat{h}^*_{\mathrm{R}}(\boldsymbol{r})$ is derived by ``copying'' the
single-particle Hamiltonian $\hat{h}_{\mathrm{R}}(\boldsymbol{r})$. The BdG
Hamiltonian can be described as 
% (i.e., $\check{H}_B = \mathrm{diag}[\hat{h}_{\mathrm{R}}, -\hat{h}^*_{\mathrm{R}} ]$). 
\begin{align}
  \check{H}_B = 
  \left[ \begin{array}{cc} 
     \hat{h}  _{\mathrm{R}} &
     \Delta_0 i \hat{\sigma}_2\\[2mm]
    -\Delta_0 i \hat{\sigma}_2&
    -\hat{h}^*_{\mathrm{R}} \\
  \end{array} \right]. 
\end{align}
In the normal state (i.e., $\Delta_0 = 0$), the energy eigenvalues of
$\check{H}_B$ are given by $ \pm {E}^{\mathrm{R}}_{\pm,n}$ as shown
schematically in Fig.~\ref{fig:LL-sche}(a). In the superconducting state,
the energy spectrum is modified by the pair potential $\Delta_0$ as shown
in Fig.~\ref{fig:LL-sche}(b), which is drawn for $\mathcal{S} = +1$.  For
the case of $\mathcal{S}=-1$, the particle (hole) branches arises in
$\mu>0$ ($\mu<0$), and hence, the sign of the slope of the chiral Majorana
mode at $\mu=0$ becomes opposite.

\bibliography{tsc03}

%\printbibliography

\end{document}